\documentclass[useAMS,usenatbib]{mn2e}
\usepackage{mn2e-breakabs}
\usepackage{graphicx}
\usepackage{subfigure}
\usepackage{times}
\usepackage{caption}
\usepackage{rotating}
\usepackage{rotfloat}
\usepackage{multicol}
\usepackage{array}
\usepackage{booktabs}
\newcommand{\Hunit}{\,{\rm km}\,{\rm s}^{-1}\,{\rm Mpc}^{-1}}

\def\la{\mathrel{\mathpalette\fun <}}
\def\ga{\mathrel{\mathpalette\fun >}}
\def\fun#1#2{\lower3.6pt\vbox{\baselineskip0pt\lineskip.9pt
        \ialign{$\mathsurround=0pt#1\hfill##\hfil$\crcr#2\crcr\sim\crcr}}}

\newcommand{\be}{\begin{equation}}
\newcommand{\ee}{\end{equation}}
\newcommand{\ba}{\begin{eqnarray}}
\newcommand{\ea}{\end{eqnarray}}
\newcommand{\simgt}{\,\hbox{\lower0.6ex\hbox{$\sim$}\llap{\raise0.6ex\hbox{$>$}}}\,}
\newcommand{\simlt}{\,\hbox{\lower0.6ex\hbox{$\sim$}\llap{\raise0.6ex\hbox{$<$}}}\,}

\begin{document}

\title[Single-Probe Cosmology Measurements from SDSS BOSS DR9 CMASS]
{The clustering of galaxies in the SDSS-III Baryon Oscillation Spectroscopic Survey:
single-probe measurements and the strong power of $f(z)\sigma_8(z)$ on constraining dark energy
}

\author[Chuang et al.]{
  \parbox{\textwidth}{
 Chia-Hsun Chuang$^1$\thanks{MultiDark Fellow; E-mail: chia-hsun.chuang@uam.es},
 Francisco Prada$^{1,2,3}$,
Antonio J. Cuesta$^4$,
Daniel J. Eisenstein$^5$, 
Eyal Kazin$^6$, 
Nikhil Padmanabhan$^4$,
Ariel G. S\'anchez$^7$, 
Xiaoying Xu$^8$,
Florian Beutler$^9$,
Marc Manera$^{10}$,
David J Schlegel$^9$,
Donald P. Schneider$^{11,12}$,
David H. Weinberg$^{13}$
Jon Brinkmann$^{14}$,
Joel R. Brownstein$^{15}$,
Daniel Thomas$^{10}$
 }
  \vspace*{4pt} \\
$^1$ Instituto de F\'{\i}sica Te\'orica, (UAM/CSIC), Universidad Aut\'onoma de Madrid,  Cantoblanco, E-28049 Madrid, Spain \\
$^2$ Campus of International Excellence UAM+CSIC, Cantoblanco, E-28049 Madrid, Spain \\
$^3$ Instituto de Astrof\'{\i}sica de Andaluc\'{\i}a (CSIC), Glorieta de la Astronom\'{\i}a, E-18080 Granada, Spain \\
$^4$ Department of Physics, Yale University, 260 Whitney Ave, New Haven, CT 06520, USA\\
$^5$ Harvard-Smithsonian Center for Astrophysics, 60 Garden St., Cambridge, MA 02138, USA\\
$^6$ Centre for Astrophysics and Supercomputing, Swinburne University of Technology, P.O. Box 218, Hawthorn, Victoria 3122, Australia\\
$^7$ Max-Planck-Institut f¨ur extraterrestrische Physik, Postfach 1312, Giessenbachstr., 85741 Garching, Germany.\\
$^8$ Department of Physics, Carnegie Mellon University, 5000 Forbes Ave., Pittsburgh, PA 15213, USA\\
$^9$ Lawrence Berkeley National Lab, 1 Cyclotron Rd, Berkeley CA 94720, USA\\
$^{10}$ Institute of Cosmology and Gravitation, University of Portsmouth, Dennis Sciama Building, Portsmouth PO1 3FX, UK\\
$^{11}$ Department of Astronomy and Astrophysics, The Pennsylvania State University, University Park, PA 16802, USA\\
$^{12}$ Institute for Gravitation and the Cosmos, The Pennsylvania State University, University Park, PA 16802, USA\\
$^{13}$ Department of Astronomy and CCAPP, Ohio State University, Columbus, OH, USA\\
$^{14}$ Apache Point Observatory, Apache Point Road, PO Box 59, Sunspot, NM 88349, USA\\
$^{15}$ Department of Physics and Astronomy, University of Utah, 115 S 1400 E, Salt Lake City, UT 84112, USA\\
}

\date{\today} 

\maketitle

\begin{abstract}
We present measurements of the anisotropic galaxy clustering from the Data Release 9 (DR9) CMASS sample of 
the SDSS-III Baryon Oscillation Spectroscopic Survey (BOSS). We analyze the broad-range shape of the monopole 
and quadrupole correlation functions to obtain constraints, at the effective redshift z=0.57 of the sample, on the Hubble expansion rate $H(z)$, 
the angular-diameter distance $D_A(z)$, the normalized growth rate $f(z)\sigma_8(z)$, 
the physical matter density $\Omega_mh^2$, and the biased amplitude of matter fluctuation $b\sigma_8(z)$. We obtain 
$\{H(0.57)$, $D_A(0.57)$, $f(0.57)\sigma_8(0.57)$, $\Omega_mh^2$, $b\sigma_8(0.57)\}$ = 
$\{87.6_{-6.8}^{+6.7}$kms$^{-1}$Mpc$^{-1}$, $1396\pm73$Mpc, $0.428\pm0.066$, $0.126_{-0.010}^{+0.008}$, $1.19\pm0.14\}$ and their covariance matrix as well.
The parameters which are not well constrained by our
galaxy clustering analysis are marginalized over with wide flat priors. 
Since no priors from other data sets (i.e., CMB) are adopted and no dark energy models are assumed, our results from BOSS CMASS galaxy clustering alone may be combined 
with other data sets, i.e. CMB, SNe, lensing or other galaxy clustering data to constrain the parameters of a given cosmological model.
We show that the major power on constraining dark energy from the anisotropic
galaxy clustering signal, as compared to the angular-averaged one (monopole),  arises
from including the normalized growth rate $f(z)\sigma_8(z)$. 
In the case of the cosmological model  
assuming a constant dark energy equation of state and a flat universe ($w$CDM),
our single-probe CMASS
constraints, 
combined with CMB (WMAP9+SPT), yield a value for the dark energy equation of state parameter of $w=-0.90\pm0.11$.
Therefore, it is important to include $f(z)\sigma_8(z)$ while investigating the nature of dark energy with current and upcoming large-scale galaxy surveys.
\end{abstract}


\section{Introduction}
The cosmic large-scale structure from galaxy redshift surveys provides a 
powerful probe of dark energy and the cosmological model
that is highly complementary to the cosmic microwave 
background (CMB) (e.g., \citealt{Hinshaw:2012fq}), supernovae (SNe) 
\citep{Riess:1998cb,Perlmutter:1998np}, and weak lensing (e.g. see \citealt{VanWaerbeke:2003uq} for a review).

The scope of galaxy redshift 
surveys has dramatically increased in the last decade. The 2dF Galaxy Redshift Survey (2dFGRS) 
obtained 221,414 galaxy redshifts at $z<0.3$ \citep{Colless:2001gk,Colless:2003wz}, 
and the Sloan Digital Sky Survey (SDSS, \citealt{York:2000gk}) has collected 
930,000 galaxy spectra in the Seventh Data Release (DR7) at $z<0.5$ \citep{Abazajian:2008wr}.
WiggleZ has collected spectra of 240,000 emission-line galaxies at $0.5<z<1$ over 
1000 square degrees \citep{Drinkwater:2009sd, Parkinson:2012vd}, and the Baryon Oscillation Spectroscopic Survey (BOSS, \citealt{Dawson13}) of the SDSS-III \citep{Eisenstein11} is surveying 1.5 million 
luminous red galaxies (LRGs) at $0.1<z<0.7$ over 10,000 square degrees.
The first BOSS data set has been made publicly available recently in SDSS data release 9 (DR9, \citealt{Ahn:2012fh}).
The planned space mission Euclid\footnote{http://sci.esa.int/euclid} will survey over 60 million emission-line galaxies at $0.7<z<2$ over 15,000 deg$^2$ 
(e.g. \citealt{RB}) and the upcoming ground-based experiment BigBOSS\footnote{http://bigboss.lbl.gov/} will survey 20 million galaxy redshifts up to $z=1.7$ and 
600,000 quasars ($2.2 < z < 3.5$) over 14,000 deg$^2$ \citep{Schelgel:2011zz}.
The WFIRST satellite would map 17 million galaxies in the redshift
range $1.3 < z < 2.7$ over 3400 deg$^2$, with a larger area 
possible with an extended mission \citep{Green:2012mj}.

Large-scale structure data from galaxy redshift surveys can be analyzed using either 
the power spectrum or the two-point correlation function. Although these two methods are 
Fourier transforms of one another, the analysis processes, the statistical uncertainties, and the systematics are quite 
different and the results cannot be converted using Fourier transform 
directly because of the finite size of the survey volume. 
The SDSS-II LRG data have been analyzed, and the cosmological results delivered, using both the power spectrum 
(see, e.g., \citealt{Tegmark:2003uf,Hutsi:2005qv,Padmanabhan:2006ku,Blake:2006kv,Percival:2007yw,Percival:2009xn,Reid:2009xm,Montesano:2011bp}), 
and the correlation function method (see, e.g., 
\citealt{Eisenstein:2005su,Okumura:2007br,Cabre:2008sz,Martinez:2008iu,Sanchez:2009jq,Kazin:2009cj,Chuang:2010dv,Samushia:2011cs,Padmanabhan:2012hf,Xu:2012fw}). 
Similar analysis have been also applied on the SDSS-III BOSS CMASS sample and obtained the most precise measurements to date 
\citep{Anderson:2012sa,Manera:2012sc,Nuza:2012mw,Reid:2012sw,Samushia:2012iq,Tojeiro:2012rp}.

Galaxy clustering allows us to differentiate smooth dark energy and modified gravity as the cause for cosmic acceleration through the simultaneous measurements 
of the cosmic expansion history $H(z)$, and the growth rate of cosmic large scale structure, $f(z)$ \citep{Guzzo08,Wang08,Blake:2012pj}. 
However, to measure $f(z)$, one must measure the galaxy bias $b$, which requires measuring higher-order statistics of the galaxy clustering (see \citealt{Verde:2001sf}).
\cite{Song09} proposed using the normalized growth rate, $f(z)\sigma_8(z)$, which would avoid the uncertainties from the galaxy bias.
\cite{Percival:2008sh} developed a method to measure $f(z)\sigma_8(z)$ and applied it on simulations.
\cite{Wang12} estimated expected statistical constraints on
dark energy and modified gravity, including redshift-space distortions and other constraints from galaxy clustering, using a Fisher matrix formalism.

In principle, the Hubble expansion rate $H(z)$, the angular-diameter distance $D_A(z)$, the normalized growth rate $f(z)\sigma_8(z)$, and
the physical matter density $\Omega_mh^2$ can be well constrained by analyzing the galaxy clustering data alone.
\cite{Eisenstein:2005su} demonstrated the feasibility of measuring $\Omega_mh^2$ and an effective distance, $D_V(z)$, 
from the SDSS DR3 LRGs, where $D_V(z)$ corresponds to a combination of $H(z)$ and $D_A(z)$. 
\cite{Chuang:2011fy} measured $H(z)$ and $D_A(z)$ simultaneously using the galaxy clustering data from 
the two dimensional two-point correlation function of SDSS DR7 LRGs.
\cite{Chuang:2012ad,Chuang:2012qt} improved the method and modeling to measure $H(z)$, $D_A(z)$, $\beta$, and $\Omega_m h^2$ from the same data.

\cite{Samushia:2011cs} measured $f(z)\sigma_8(z)$ from the SDSS DR7 LRGs.
\cite{Blake:2012pj} measured $H(z)$, $D_A(z)$, and $f(z)\sigma_8(z)$ from the WiggleZ Dark Energy Survey galaxy sample.
\cite{Reid:2012sw} measured $H(z)$, $D_A(z)$, and $f(z)\sigma_8(z)$ from the SDSS BOSS DR9 CMASS
and \cite{Samushia:2012iq} derived the cosmological implications from these measurements to test deviations from the concordance $\Lambda$CDM model and general relativity 
(see also \cite{Nesseris:2011pc} for using $f(z)\sigma_8(z)$ to constrain modified gravity theories).

In this study, we apply the similar method and model as \cite{Chuang:2012ad,Chuang:2012qt} to measure $H(z)$, $D_A(z)$, $f(z)\sigma_8(z)$, and $\Omega_m h^2$ 
which extracts a summary of the cosmological information from the large-scale structure of the SDSS BOSS DR9 CMASS alone. 
One can combine our single-probe measurements with other data sets (i.e. CMB, SNe, etc.) to constrain the cosmological parameters of a given dark energy model.  
We also explore the strong power of adding $f(z)\sigma_8(z)$ to the two dimensional galaxy clustering analysis on constraining dark energy. 

This study is part of a series of papers performing anisotropic clustering analysis on the BOSS DR9 CMASS galaxy sample.
Instead of using multipoles, \cite{Sanchez13} present a different method taken from the 'clustering wedges' measurements by \cite{Kazin13}
and combine the results with CMB, SNe, and Baryon Acoustic Oscillations (BAO) to obtain constraints on the cosmological parameters.
\cite{Anderson13} present anisotropic analysis using two approaches, multipoles and wedges, to obtain robust measurement of the BAO signal.

This paper is organized as follows. In Section \ref{sec:data}, we introduce the SDSS-III/BOSS DR9 galaxy sample and mock catalogues used 
in our study. In Section \ref{sec:method}, we describe the details of the 
methodology that constrains cosmological parameters from our galaxy clustering analysis. 
In Section \ref{sec:results}, we present our single-probe cosmological measurements and demonstrate how 
to use our results assuming different cosmological models or combining other data sets. 
In Section \ref{sec:compare}, we compare our results with previous or parallel works.
In Section \ref{sec:single-probe}, we discuss the requirements to provide single-probe measurements.
In Section \ref{sec:test}, we apply some systematic tests to our measurements.
We summarize and conclude in Sec.~\ref{sec:conclusion}.

\section{Data Set} \label{sec:data}
\subsection{The CMASS Galaxy Sample}
\label{sec:cmass}
The Sloan Digital Sky Survey (SDSS; \citealt{Fukugita:1996qt,Gunn:1998vh,York:2000gk,Smee:2012wd}) mapped over one quarter 
of the sky using the dedicated 2.5m Sloan Telescope \citep{Ahn:2012fh}.
The Baryon Oscillation Sky Survey (BOSS, \citealt{Eisenstein11, Bolton:2012hz, Dawson13}) is part of the SDSS-III survey. 
It is collecting the spectra and redshifts for 1.5 million galaxies, 160,000 quasars and 
100,000 ancillary targets. The Data Release 9 has been made publicly available\footnote{http://www.sdss3.org/}.
We use galaxies from the SDSS-III BOSS DR9 CMASS catalogue in the redshift range $0.43<z<0.7$. 
'CMASS' samples are selected with an approximately constant stellar mass threshold \citep{Eisenstein11}.
The sample we are using includes a total of 264,283 galaxies with 207,246 in the north and 57,037 in the 
south Galactic hemispheres. The median redshift of the sample is $z=0.57$.
The details of generating this sample are described in \cite{Dawson13}.

\subsection{The Mock Catalogues}
\cite{Manera:2012sc} created 600 mock catalogues for DR9 CMASS sample. They created 2nd-order Lagrangian perturbation theory 
matter fields from which they populate haloes with mock galaxies using a halo occupation distribution prescription 
which has been calibrated to reproduce the clustering measurements on scales between 30 and 80 $h^{-1}$Mpc \citep{White:2010ed}. We use
these mock catalogues to construct the covariance matrix in our analysis.

\section{Methodology} 
\label{sec:method}

In this section, we describe the measurement of the multipoles of the correlation function
from the observational data, construction of the theoretical prediction, 
and the likelihood analysis that leads to constraining 
cosmological parameters and dark energy.

\subsection{Measuring the Two-Dimensional Two-Point Correlation Function}

We convert the measured redshifts of the BOSS CMASS galaxies to comoving distances 
by assuming a fiducial model, i.e., flat $\Lambda$CDM with $\Omega_m=0.274$ and $h=0.7$ which is the same model for constructing the mock catalogues (see \citealt{Manera:2012sc}). 
We use the two-point correlation function estimator given by 
\cite{Landy:1993yu}:
\begin{equation}
\label{eq:xi_Landy}
\xi(\sigma,\pi) = \frac{DD(\sigma,\pi)-2DR(\sigma,\pi)+RR(\sigma,\pi)}{RR(\sigma,\pi)},
\end{equation}
where $\pi$ is the separation along the light of sight (LOS) and $\sigma$ 
is the separation in the plane of the sky. DD, DR, and RR represent the normalized 
data-data,
data-random, and random-random pair counts, respectively, for a given
distance range. The LOS is defined as the direction from the observer to the 
center of a galaxy pair. The bin size we use is
$1 \, h^{-1}$Mpc$\times 1 \,h^{-1}$Mpc. 
The Landy and Szalay estimator has minimal variance for a Poisson
process. Random data are generated with the same radial
and angular selection functions as the real data. One can reduce the shot noise due
to random data by increasing the amount of random data. The number
of random data we use is more than 15 times that of the real data. While
calculating the pair counts, we assign to each data point a radial
weight of $1/[1+n(z)\cdot P_w]$, where $n(z)$ is the radial
number density and $P_w = 2\cdot 10^4$ $h^{-3}$Mpc$^3$ (see  
\citealt{Feldman:1993ky}).

\subsection{Theoretical Two-Dimensional Two-Point Correlation Function}

First, we adopt the cold dark matter model and the simplest inflation model (adiabatic initial condition).
Thus, we can compute the linear matter power spectra, $P_{lin}(k)$, by using CAMB (Code for Anisotropies in the Microwave Background, \citealt{Lewis:1999bs}). The linear power spectrum can be decomposed into two parts:
\begin{equation} \label{eq:pk_lin}
P_{lin}(k)=P_{nw}(k)+P_{BAO}^{lin}(k),
\end{equation}
where $P_{nw}(k)$ is the ``no-wiggle'' or pure CDM power spectrum calculated using Eq.(29) from \cite{Eisenstein:1997ik}. $P_{BAO}^{lin}(k)$ is the wiggled part defined by the equation itself.
The nonlinear damping effect of the ``wiggled'' part, in redshift space, can be well approximated following \cite{Eisenstein:2006nj} by
\begin{equation} \label{eq:bao}
P_{BAO}^{nl}(k,\mu_k)=P_{BAO}^{lin}(k)\cdot \exp\left(-\frac{k^2}{2k_\star^2}[1+\mu_k^2(2f+f^2)]\right),
\end{equation}
where $\mu_k$ is the cosine of the angle between ${\bf k}$ and the LOS, $f$ is the growth rate, and
$k_\star$ is computed following \cite{Crocce:2005xz, Matsubara:2007wj} by
\begin{equation} \label{eq:kstar}
k_\star=\left[\frac{1}{3\pi^2}\int P_{lin}(k)dk\right]^{-1/2}.
\end{equation}
The dewiggled power spectrum is
\begin{equation} \label{eq:pk_dw}
P_{dw}(k,\mu_k)=P_{nw}(k)+P_{BAO}^{nl}(k,\mu_k).
\end{equation}

Next, we include the linear redshift distortion as follows in order to
obtain the galaxy power spectrum in redshift space at large scales \citep{Kaiser:1987qv}, i.e.,
\begin{eqnarray} \label{eq:pk_2d}
P_g^s(k,\mu_k)&=&b^2(1+\beta\mu_k^2)^2P_{dw}(k,\mu_k),
\end{eqnarray}
where $b$ is the linear galaxy bias and $\beta$ is the linear redshift distortion parameter.

We compute the theoretical two-point correlation 
function, $\xi^\star(\sigma,\pi)$, by Fourier transforming the non-linear power spectrum
$P_g^s(k,\mu_k)$. This task is efficiently performed by using Legendre polynomial expansions and one-dimensional integral convolutions as introduced in \cite{Chuang:2012qt}.

We convolve the 2D correlation function with the distribution function of 
random pairwise velocities, $f_v(v)$, to obtain the final model $\xi(\sigma,\pi)$ following
\cite{peebles1980} by
\begin{equation} \label{eq:theory}
 \xi(\sigma,\pi)=\int_{-\infty}^\infty \xi^\star\left(\sigma,\pi-\frac{v}{H(z)a(z)}
 \right)\,f_v(v)dv,
\end{equation}
where the random motions (fingers of god) are represented by an exponential form 
(e.g., \citealt{Ratcliffe:1997kz,Landy:2002xg})
\begin{equation}
 f_v(v)=\frac{1}{\sigma_v\sqrt{2}}\exp\left(-\frac{\sqrt{2}|v|}{\sigma_v}\right),
\end{equation}
where $\sigma_v$ is the pairwise peculiar velocity dispersion.

The cosmological parameter set that we use to compute the theoretical
correlation function is 
$\{H(z)$, $D_A(z)$, $\beta(z)$, $\Omega_mh^2$, $b\sigma_8(z)$, $\Omega_bh^2$, $n_s$, $\sigma_v, f(z)\}$, where
$\Omega_m$ and $\Omega_b$ are the matter and
baryon density fractions, $n_s$ is the power-law index of the primordial matter power spectrum, 
$h$ is the dimensionless Hubble
constant ($H_0=100h$ km s$^{-1}$Mpc$^{-1}$), and $\sigma_8(z)$ is the normalization of the power spectrum.
The linear redshift distortion parameter can be expressed as $\beta(z)=f(z)/b$.
Thus, one can derive $f(z)\sigma_8(z)$ from the measured $\beta(z)$ and $b\sigma_8(z)$.
On the scales we use for comparison with the BOSS CMASS data, the theoretical correlation 
function only depends on cosmic curvature and dark energy through the
parameters $H(z)$, $D_A(z)$, $\beta(z)$, and $b\sigma_8(z)$        
assuming that dark energy perturbations are unimportant (valid in the simplest dark energy models).
Thus we are able to extract constraints from clustering data that are independent of a dark energy 
model and cosmic curvature.

\subsection{Effective Multipoles of the Correlation Function}  \label{sec:multipoles}

The traditional multipoles of the two-point correlation function, in redshift space, are defined by
\ba
\label{eq:multipole_1}
\xi_l(s) &\equiv & \frac{2l+1}{2}\int_{-1}^{1}{\rm d}\mu\, \xi(\sigma,\pi) P_l(\mu)\nonumber\\
&=& \frac{2l+1}{2}\int_{0}^{\pi}{\rm d}\theta \, \sqrt{1-\mu^2}\, \xi(\sigma,\pi) P_l(\mu),
\ea
where 
\begin{eqnarray}
\mu&\equiv&\frac{\pi}{\sqrt{\sigma^2+\pi^2}},\\
\theta&\equiv&\cos^{-1}\mu, 
\end{eqnarray}
and $P_l(\mu)$ is the Legendre Polynomial ($l=$0 and 2 here). 
We integrate over a spherical shell with radius $s$,
while actual measurements of $\xi(\sigma,\pi)$ are done in discrete bins.
To compare the measured $\xi(\sigma,\pi)$ and our theoretical model, the last integral in Eq.(\ref{eq:multipole_1}) should be converted into a sum.
This leads to the definition for the effective multipoles of the correlation function \citep{Chuang:2012ad}:
\begin{equation}\label{eq:multipole}
 \hat{\xi}_l(s) \equiv \frac{\displaystyle\sum_{s-\frac{\Delta s}{2} < \sqrt{\sigma^2+\pi^2} < s+\frac{\Delta s}{2}}(2l+1)\xi(\sigma,\pi)P_l(\mu)\sqrt{1-\mu^2}}{\mbox{Number of bins used in the numerator}},
\end{equation}
where $\Delta s=5$ $h^{-1}$Mpc in this work, and 
\begin{equation}
\sigma=(n+\frac{1}{2})\mbox{$h^{-1}$Mpc}, n=0,1,2,...
\end{equation}
\begin{equation}
\pi=(m+\frac{1}{2})\mbox{$h^{-1}$Mpc}, m=0,1,2,...
\end{equation}

Both the measurement and the theoretical prediction for the effective multipoles are computed using
Eq.(\ref{eq:multipole}), with $\xi(\sigma,\pi)$ given by the measured correlation function
(see Eq.\ref{eq:xi_Landy}) for the measured effective multipoles, and 
Eq.(\ref{eq:theory}) for the theoretical predictions.
We do not use the conventional definitions of multipoles to extract parameter constraints
as they use continuous integrals (see Eq. \ref{eq:multipole_1}). 
Bias of the result could be introduced if the definitions of multipoles differ between measurements from
data and the theoretical model.

\subsection{Covariance Matrix} \label{sec:covar}

We use the 600 mock catalogues created by \cite{Manera:2012sc} for the BOSS CMASS DR9
to estimate the covariance matrix of the observed correlation function. 
We calculate the multipoles of the correlation functions 
of the mock catalogues and construct the covariance matrix as
\begin{equation}
 C_{ij}=\frac{1}{N-1}\sum^N_{k=1}(\bar{X}_i-X_i^k)(\bar{X}_j-X_j^k),
\label{eq:covmat}
\end{equation}
where $N$ is the number of the mock catalogues, $\bar{X}_m$ is the
mean of the $m^{th}$ element of the vector from the mock catalogue multipoles, and
$X_m^k$ is the value in the $m^{th}$ elements of the vector from the $k^{th}$ mock
catalogue multipoles. The data vector ${\bf X}$ is defined by Eq.(\ref{eq:X}).

\subsection{Likelihood}
The likelihood is taken to be proportional to $\exp(-\chi^2/2)$ \citep{press92}, 
with $\chi^2$ given by
\begin{equation} \label{eq:chi2}
 \chi^2\equiv\sum_{i,j=1}^{N_{X}}\left[X_{th,i}-X_{obs,i}\right]
 C_{ij}^{-1}
 \left[X_{th,j}-X_{obs,j}\right]
\end{equation}
where $N_{X}$ is the length of the vector used, 
$X_{th}$ is the vector from the theoretical model, and $X_{obs}$ 
is the vector from the observed data.

As explained in \cite{Chuang:2011fy}, instead of recalculating the observed correlation function while 
computing for different models, we rescale the theoretical correlation function to avoid rendering the $\chi^2$ values arbitrary.
It can be considered as an application of Alcock-Paczynski effect \citep{Alcock:1979mp}.
The rescaled theoretical correlation function is computed by
\begin{equation} \label{eq:inverse_theory_2d}
 T^{-1}(\xi_{th}(\sigma,\pi))=\xi_{th}
 \left(\frac{D_A(z)}{D_A^{fid}(z)}\sigma,
 \frac{H^{fid}(z)}{H(z)}\pi\right),
\end{equation}
where $\xi_{th}$ is computed by eq. (\ref{eq:theory}), and $\chi^2$ can be rewritten as
\ba 
\label{eq:chi2_2}
\chi^2 &\equiv&\sum_{i,j=1}^{N_{X}}
 \left\{T^{-1}X_{th,i}-X^{fid}_{obs,i}\right\}
 C_{fid,ij}^{-1} \cdot \nonumber\\
 & & \cdot \left\{T^{-1}X_{th,j}-X_{obs,j}^{fid}\right\};
\ea
where $T^{-1}X_{th}$ is the vector computed by eq.\ (\ref{eq:multipole}) from the rescaled theoretical correlation function, eq. (\ref{eq:inverse_theory_2d}).
$X^{fid}_{obs}$ is the vector from observed data measured with the fiducial model (see \citealt{Chuang:2011fy} for more details regarding the rescaling method).

\subsection{Markov Chain Monte-Carlo Likelihood Analysis} \label{sec:mcmc}

We perform Markov Chain Monte-Carlo
likelihood analyses using CosmoMC \citep{Lewis:2002ah}. 
The parameter space that we explore spans the parameter set of
$\{H(0.57)$, $D_A(0.57)$, $\Omega_mh^2$, $\beta(0.57)$, $b\sigma_8(0.57)$, $\Omega_bh^2$, $n_s$, $\sigma_v$, $f(0.57)\}$. 
Only $\{H(0.57)$, $D_A(0.57)$, $\Omega_mh^2$, $\beta(0.57)$, $b\sigma_8(0.57)\}$ are well constrained using
the BOSS CMASS alone in the scale range of interest. We marginalize over the other parameters, 
$\{\Omega_bh^2$, $n_s$, $\sigma_v$, $f(0.57)\}$, with the flat priors 
$\{(0.01859,0.02657)$, $(0.865,1.059)$, $(0,500)$ km/s, $(0.5,1)\}$, 
where the flat priors of $\Omega_b h^2$ and $n_s$ are centered on 
the WMAP7 measurements with a width of $\pm7\sigma_{WMAP}$ ($\sigma_{WMAP}$ is taken from
\citealt{Komatsu:2010fb}). These priors
are sufficiently wide to ensure that CMB constraints are not double counted 
when our results are combined with CMB data \citep{Chuang:2010dv}.

\section{Results} \label{sec:results}

\subsection{Measurement of multipoles}

\begin{figure*}
\begin{center}
 \subfigure{\label{fig:mono}\includegraphics[width=1 \columnwidth,clip,angle=0]{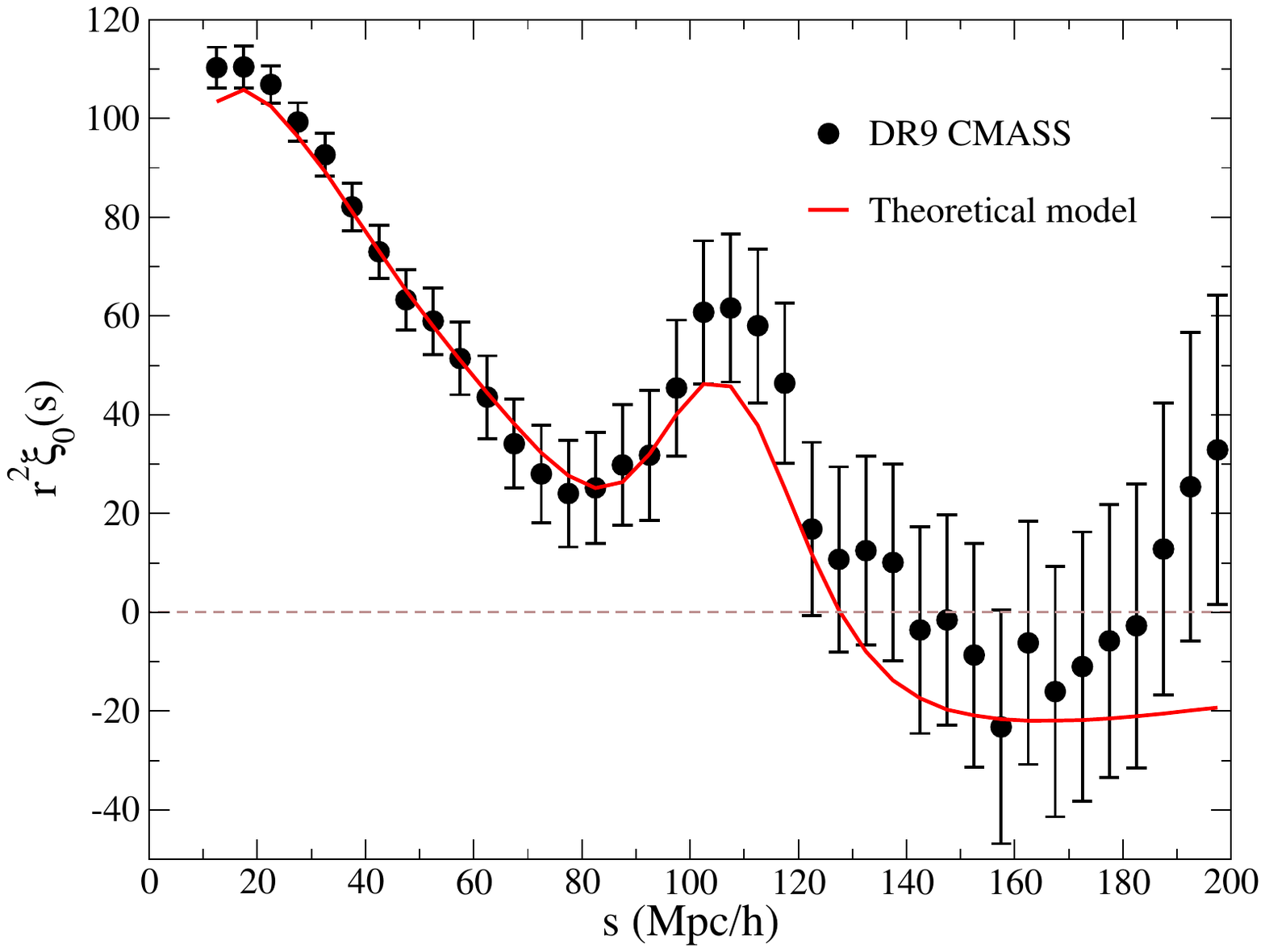}}
 \subfigure{\label{fig:quad}\includegraphics[width=1 \columnwidth,clip,angle=0]{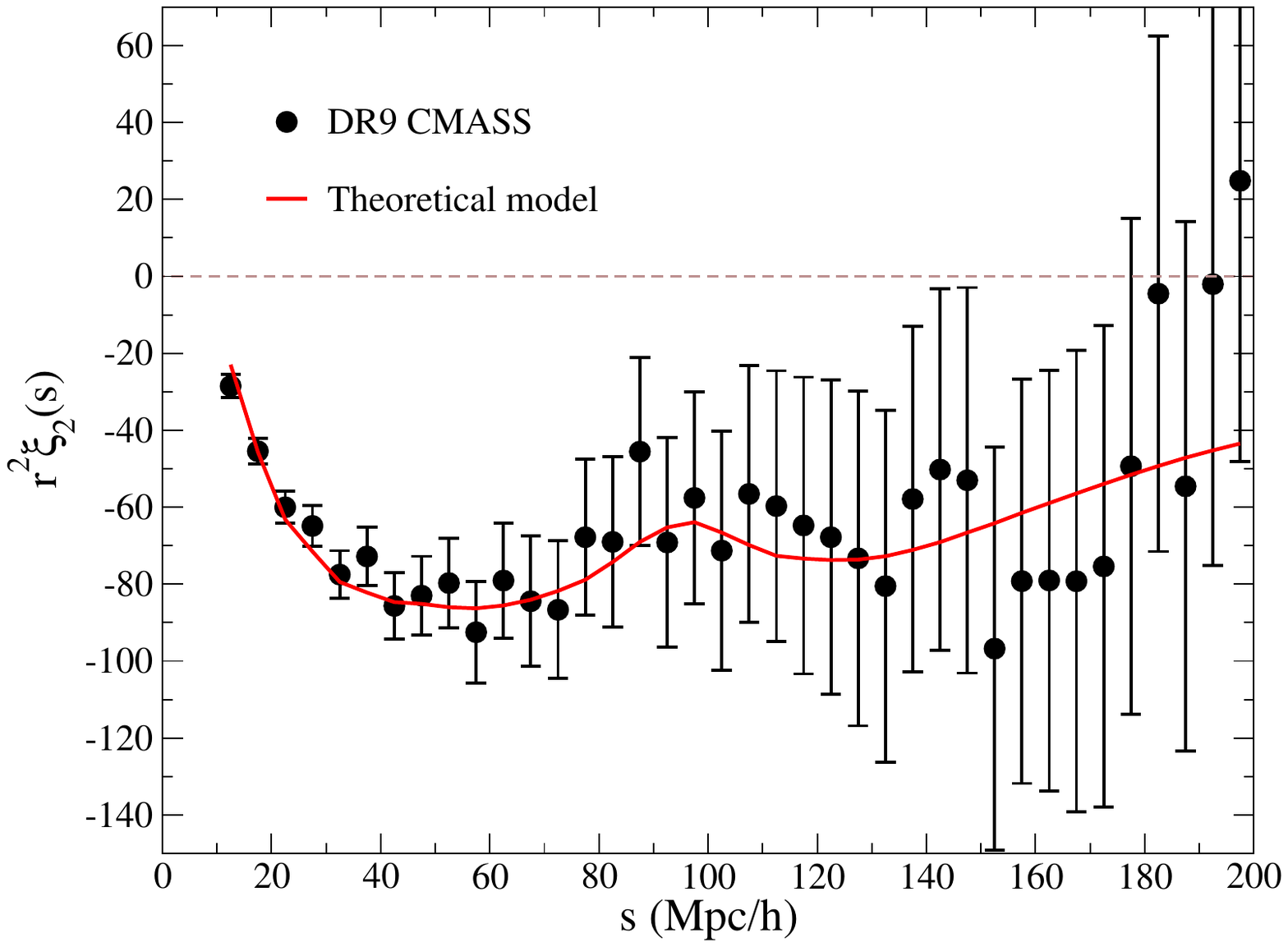}}
\end{center}
\caption{
Measurement of effective monopole (left) and quadrupole (right) of the correlation function for the BOSS DR9 CMASS galaxy sample (black points), compared to the the theoretical model 
given the parameters measured (solid line). The error bars are the square roots of the diagonal elements of the covariance matrix (see Sec. \ref{sec:covar}).
In this study, our fitting scale range is $40h^{-1}$Mpc $<s<160h^{-1}$Mpc.
}
\label{fig:multipoles}
\end{figure*}

Fig.\ref{fig:mono} and \ref{fig:quad} show the effective 
monopole ($\hat{\xi}_0$) and quadrupole ($\hat{\xi}_2$) measured from the BOSS CMASS galaxy sample
compared with the theoretical model given the parameters measured. 
We are using the scale range, $s=40-160\,h^{-1}$Mpc, and the bin size is 5 $h^{-1}$Mpc. 
The data points from the multipoles in the scale range considered are combined to form a 
vector, $X$, i.e.,
\be
{\bf X}=\{\hat{\xi}_0^{(1)}, \hat{\xi}_0^{(2)}, ..., \hat{\xi}_0^{(N)}; 
\hat{\xi}_2^{(1)}, \hat{\xi}_2^{(2)}, ..., \hat{\xi}_2^{(N)};...\},
\label{eq:X}
\ee
where $N$ is the number of data points in each measured multipole; here $N=24$.
The length of the data vector ${\bf X}$ depends on the number of multipoles used. 

\subsection{Measurement of Cosmological Parameters from BOSS CMASS only}
\label{sec:CMASS_result}

We now present the dark energy model independent measurements of the parameters
$\{H(0.57)$, $D_A(0.57)$, $\Omega_m h^2$, $\beta(0.57)$, and b$\sigma_8(0.57)\}$, obtained by using the 
method described in previous sections. We also present the derived 
parameters including $H(0.57)\,r_s(z_d)$, $D_A(0.57)/r_s(z_d)$, $f(0.57)\sigma_8(0.57)$,
$D_V(0.57)/r_s(z_d)$, $A_s(0.57)$, $\alpha$, and $\epsilon$ 
\begin{equation} \label{eq:dv}
 D_V(z)\equiv \left[(1+z)^2D_A(z)^2\frac{cz}{H(z)}\right]^\frac{1}{3},
\end{equation} 
\begin{equation}\label{eq:A}
 A_s(z)\equiv D_V(z)\frac{\sqrt{\Omega_mH_0^2}}{cz},
\end{equation}

\begin{equation}\label{eq:alpha}
 \alpha\equiv \frac{D_V(z)/r_s(z_d)}{(D_V(z)/r_s(z_d))_{fid}},
\end{equation}
and
\begin{equation}\label{eq:epsilon}
 \epsilon\equiv \left[\frac{(H(z)D_A(z))_{fid}}{H(z)D_A(z)}\right]^{1/3}-1,
\end{equation}
where $r_s(z_d)$ is the comoving sound horizon at the drag epoch calculated
using eq. (6) in \cite{Eisenstein:1997ik}.
$D_V(z)$ is the effective distance which can be measured from the spherical averaged correlation function or power spectrum (e.g. see \citealt{Eisenstein:2005su}).
$A_s(z)$ is a robust measurement while including small scales (e.g. see \citealt{Blake:2012pj}).
$\alpha$ and $\epsilon$ are the dilation and wrapping parameters between the true and fiducial cosmology models (e.g. see \citealt{Xu:2012fw}).

Table \ref{table:mean_fid} lists the mean, rms variance, and 68\%
confidence level limits for $\{H(0.57)$, $D_A(0.57)$, $\Omega_m h^2$, 
$\beta(0.57)$, b$\sigma_8(0.57)$, $H(0.57) \,r_s(z_d)/c$, $D_A(0.57)/r_s(z_d)$, $D_V(0.57)/r_s(z_d)$, 
$A_s(0.57)$, $\alpha$, and $\epsilon\}$ derived in an MCMC likelihood analysis from the measured $\hat{\xi}_0+\hat{\xi}_2$ 
of the DR9 CMASS correlation function.

Table \ref{table:covar_matrix_fid} gives the normalized covariance matrix
for this parameter set measured using $\hat{\xi}_0+\hat{\xi}_2$.
It is clear that the correlation between $\Omega_m h^2$ and $D_V(0.57)/r_s(z_d)$, $D_A(0.57)/r_s(z_d)$, or $r_s(z_d) H(0.57)/c$ are close to zero, since the dependency on $\Omega_m h^2$ is removed by dividing or multiplying $r_s(z_d)$.

For this measurement, we use 48 bins ($\hat{\xi}_0+\hat{\xi}_2$), 9 fitting parameters (see Sec. \ref{sec:mcmc}), 
and scale range $40h^{-1}$Mpc $<s<160h^{-1}$Mpc. The $\chi^2$ per degree of freedom (d.o.f.) is 0.51. 
This low value might indicate not only a good modeling but also possible over-estimation for the covariance matrix constructed with the mock catalogues.

\begin{table}
\begin{center}
\begin{tabular}{crrrr}\hline
Measured &mean &$\sigma$ &lower &upper \\ \hline
$	H(0.57)	 $ &\ \ 	87.6	&\ \ 	7.2	&\ \ 	80.8	&\ \ 	94.3	\\
$	D_A(0.57)$ &\ \ 		1396	&\ \ 	74	&\ \ 	1324	&\ \ 	1470	\\
$	\Omega_m h^2	$&\ \	0.126	&\ \ 	0.019	&\ \ 	0.116	&\ \ 	0.134	\\
$	\beta=f(0.57)/b$&\ \ 	0.367	&\ \ 	0.084	&\ \ 	0.287	&\ \ 	0.446	\\
$	b\sigma_8(0.57)	$&\ \ 	1.19	&\ \ 	0.14	&\ \ 	1.05	&\ \ 	1.33	\\
\hline
Derived\\
\hline
$	r_s(z_d)H(0.57)/c	$&\ \ 	0.0454	&\ \ 	0.0031	&\ \ 	0.0426	&\ \ 	0.0482	\\
$	D_A(0.57)/r_s(z_d)	$&\ \ 	8.95	&\ \ 	0.27	&\ \ 	8.69	&\ \ 	9.22	\\
$	D_V(0.57)/r_s(z_d)	$&\ \ 	13.54	&\ \ 	0.29	&\ \ 	13.26	&\ \ 	13.82	\\
$	f(0.57)\sigma_8(0.57)	$&\ \ 	0.428	&\ \ 	0.069	&\ \ 	0.362	&\ \ 	0.494	\\
$	A_s(0.57)	$&\ \ 	0.436	&\ \ 	0.017	&\ \ 	0.419	&\ \ 	0.454	\\
$	\alpha	$&\ \ 	1.024	&\ \ 	0.022	&\ \ 	1.002	&\ \ 	1.045	\\
$	\epsilon	$&\ \ 	0.015	&\ \ 	0.029	&\ \ 	-0.012	&\ \ 	0.042	\\
\hline
\end{tabular}
\end{center}
\caption{
The mean, standard deviation, and the 68\% C.L. bounds of the measured parameters
and the derived parameters 
from the BOSS DR9 CMASS galaxy clustering. The quantity $\alpha$ is defined as $\frac{D_V(0.57)}{r_s(z_d)}/\frac{D_V(0.57)_{fid}}{r_s(z_d)_{fid}}$.
The unit of $H$ is $\Hunit$. The unit of $D_A$ and $r_s(z_d)$ is $\rm Mpc$.
} \label{table:mean_fid}
\end{table}

\begin{table*}
\centering
\begin{tabular}{crrrrrrrrrrrr}\hline
		&$H$ &$D_A$ &$\Omega_m h^2$ &$\beta$ &$b\sigma_8$ &$r_s H/c$ \\ \hline
$	H(0.57)	$&\ \ 	1.0000	&\ \ 	-0.2203	&\ \ 	0.5224	&\ \ 	0.1701	&\ \ 	0.1534	&\ \ 	0.8502\\
$	D_A(0.57)$&\ \ 		-0.2203	&\ \ 	1.0000	&\ \ 	-0.7799	&\ \ 	0.4611	&\ \ 	-0.2938	&\ \ 	0.2530\\
$	\Omega_m h^2	$&\ \	0.5224	&\ \ 	-0.7799	&\ \ 	1.0000	&\ \ 	-0.4100	&\ \ 	0.5746	&\ \ 	0.0237\\
$	\beta=f(0.57)/b	$&\ \ 	0.1701	&\ \ 	0.4611	&\ \ 	-0.4100	&\ \ 	1.0000	&\ \ 	-0.7220	&\ \ 	0.4219\\
$	b\sigma_8(0.57)	$&\ \ 	0.1534	&\ \ 	-0.2938	&\ \ 	0.5746	&\ \ 	-0.7220	&\ \ 	1.0000	&\ \ 	-0.0952\\
$	r_s(z_d)\ H(0.57)/c	$&\ \ 	0.8502	&\ \ 	0.2530	&\ \ 	0.0237	&\ \ 	0.4219	&\ \ 	-0.0952	&\ \ 	1.0000\\
$	D_A(0.57)/r_s(z_d)	$&\ \ 	0.4038	&\ \ 	0.5739	&\ \ 	-0.0033	&\ \ 	0.3153	&\ \ 	0.1183	&\ \ 	0.4874\\
$	D_V(0.57)/r_s(z_d)	$&\ \ 	-0.5020	&\ \ 	0.2651	&\ \ 	-0.0200	&\ \ 	-0.1470	&\ \ 	0.2124	&\ \ 	-0.5828\\
$	f(0.57)\sigma_8(0.57)	$&\ \ 	0.3565	&\ \ 	0.4387	&\ \ 	-0.1612	&\ \ 	0.8495	&\ \ 	-0.2768	&\ \ 	0.5300\\
$	A_s(0.57)	$&\ \ 	0.0802	&\ \ 	-0.3945	&\ \ 	0.7669	&\ \ 	-0.4649	&\ \ 	0.6870	&\ \ 	-0.3090\\
$	\alpha	$&\ \ 	-0.5020	&\ \ 	0.2651	&\ \ 	-0.0200	&\ \ 	-0.1470	&\ \ 	0.2124	&\ \ 	-0.5828\\
$	\epsilon	$&\ \ 	-0.7989	&\ \ 	-0.4034	&\ \ 	-0.0102	&\ \ 	-0.4397	&\ \ 	0.0348	&\ \ 	-0.9476\\
\hline
\end{tabular}
\begin{tabular}{crrrrrrrrrrrr}\hline
		&$D_A/r_s$ &$D_V/r_s$ &$	f\sigma_8$ &$A_s$ &$\alpha$ &$\epsilon	$ \\ \hline
$	H(0.57)	$&\ \ 	0.4038	&\ \ 	-0.5020	&\ \ 	0.3565	&\ \ 	0.0802	&\ \ 	-0.5020	&\ \ 	-0.7989	\\
$	D_A(0.57)$&\ \ 		0.5739	&\ \ 	0.2651	&\ \ 	0.4387	&\ \ 	-0.3945	&\ \ 	0.2651	&\ \ 	-0.4034	\\
$	\Omega_m h^2	$&\ \	-0.0033	&\ \ 	-0.0200	&\ \ 	-0.1612	&\ \ 	0.7669	&\ \ 	-0.0200	&\ \ 	-0.0102	\\
$	\beta=f(0.57)/b	$&\ \ 	0.3153	&\ \ 	-0.1470	&\ \ 	0.8495	&\ \ 	-0.4649	&\ \ 	-0.1470	&\ \ 	-0.4397	\\
$	b\sigma_8(0.57)	$&\ \ 	0.1183	&\ \ 	0.2124	&\ \ 	-0.2768	&\ \ 	0.6870	&\ \ 	0.2124	&\ \ 	0.0348	\\
$	r_s(z_d)H(0.57)/c	$&\ \ 	0.4874	&\ \ 	-0.5828	&\ \ 	0.5300	&\ \ 	-0.3090	&\ \ 	-0.5828	&\ \ 	-0.9476	\\
$	D_A(0.57)/r_s(z_d)	$&\ \ 	1.0000	&\ \ 	0.4220	&\ \ 	0.5389	&\ \ 	0.2301	&\ \ 	0.4220	&\ \ 	-0.7316	\\
$	D_V(0.57)/r_s(z_d)	$&\ \ 	0.4220	&\ \ 	1.0000	&\ \ 	-0.0515	&\ \ 	0.5431	&\ \ 	1.0000	&\ \ 	0.3081	\\
$	f(0.57)\sigma_8(0.57)	$&\ \ 	0.5389	&\ \ 	-0.0515	&\ \ 	1.0000	&\ \ 	-0.1512	&\ \ 	-0.0515	&\ \ 	-0.6026	\\
$	A_s(0.57)	$&\ \ 	0.2301	&\ \ 	0.5431	&\ \ 	-0.1512	&\ \ 	1.0000	&\ \ 	0.5431	&\ \ 	0.1678	\\
$	\alpha	$&\ \ 	0.4220	&\ \ 	1.0000	&\ \ 	-0.0515	&\ \ 	0.5431	&\ \ 	1.0000	&\ \ 	0.3081	\\
$	\epsilon	$&\ \ 	-0.7316	&\ \ 	0.3081	&\ \ 	-0.6026	&\ \ 	0.1678	&\ \ 	0.3081	&\ \ 	1.0000	\\
\hline
\end{tabular}
 \caption{Normalized covariance matrix of the measured and derived parameters for the BOSS CMASS DR9 galaxy clustering.} 
 \label{table:covar_matrix_fid}
\end{table*}

\subsection{Using Our Results from CMASS only}
In this section, we describe the steps to combine our results with other data sets assuming some dark energy models. 
For a given model and cosmological parameters, including the linear galaxy bias $b$, one can compute 
$H(0.57)$, $D_A(0.57)$, $\Omega_m h^2$, $b\sigma_8(0.57)$, and $f(0.57)\sigma_8(0.57)$. From Table \ref{table:mean_fid} and \ref{table:covar_matrix_fid},
one can derive the covariance matrix, $M_{ij}$, of these five parameters. Then, $\chi^2$ can be computed by
\begin{equation}
 \chi^2=\Delta_{CMASS}M_{ij}^{-1}\Delta_{CMASS},
\end{equation}
where 
\begingroup
\everymath{\scriptstyle}
\small
\begin{equation}
 \Delta_{CMASS}=\left(\begin{array}{c}H(0.57)-87.6 \\ D_A(0.57)-1396 \\ \Omega_mh^2-0.126 \\ b\sigma_8(0.57)-1.19 \\f(0.57)\sigma_8(0.57)-0.428\end{array}\right)
\end{equation}
\endgroup
and
\begingroup
\everymath{\scriptstyle}
\small
\begin{equation}
M_{ij}^{-1}=\left(\begin{array}{ccccc}
0.03850& -0.001141& -13.53& 0.4007& -1.271\\
-0.001141& 0.0008662& 3.354& -0.1598& -0.3059\\
-13.53& 3.354& 19370& -987.8& -770.0\\ 
0.4007&-0.1598& -987.8&110.8& 78.67\\ 
-1.271& -0.3059& -770.0& 78.67& 411.3
\end{array}\right).
\end{equation}
\endgroup

One can use a subset of these parameters (measured and derived) and their covariance matrix to derive the cosmological parameters. 
For example, if one is only interested in the cosmological parameters but not in the galaxy bias, $b$, 
one can use only four parameters, $H(0.57)$, $D_A(0.57)$, $\Omega_m h^2$, and $f(0.57)\sigma_8(0.57)$ to compute $\chi^2$ to constrain the parameters of a given model. 
In Sec. \ref{sec:fsigma8}, we use $D_V(0.57)/r_s(z_d)$, $\{H(0.57) r_s(z_d), D_A(0.57)/r_s(z_d)\}$, 
and $\{H(0.57) r_s(z_d), D_A(0.57)/r_s(z_d), f(0.57)\sigma_8(0.57)\}$ to explore the power on constraining dark energy from $f(0.57)\sigma_8(0.57)$.

In addition, we use $H(z)$, $D_A(z)$, and $\Omega_m h^2$ instead of $H(z)r_s(z_d)$, $D_A(z)/r_s(z_d)$ to be more general. 
For example, while combining the supernovae data, which do not have $\Omega_b h^2$ as a parameter of the cosmological model, it is simpler to use $H(z)$ than use $H(z)r_s(z_d)$.

We also provide the code for using CosmoMC that includes BOSS CMASS clustering alone\footnote{http://members.ift.uam-csic.es/chuang/BOSSDR9singleprobe}.

\subsection{Assuming Dark Energy Models} 
\label{sec:models}
In this section, we present examples of combining our CMASS-only clustering results with CMB data sets assuming specific dark energy models. 

Table \ref{table:lcdm}, \ref{table:olcdm}, \ref{table:wcdm}, and \ref{table:owcdm} show the cosmological constraints assuming 
$\Lambda$CDM, o$\Lambda$CDM (non-flat $\Lambda$CDM), $w$CDM (constant equation of state of dark energy), and o$w$CDM (non-flat universe with a constant equation of state of dark energy) models. 
In this study, we only list the parameters which can be well constrained by galaxy clustering. 
We also present the results of the combination of CMASS and CMB data.
The CMB data we use includes 
WMAP7 and WMAP9, which are the previous and the newest data release from the Wilkinson Microwave Anisotropy Probe collaboration; 
\citep{Komatsu:2010fb,Bennett:2012fp,Hinshaw:2012fq}.
We are also using the newest data release from the South Pole Telescope (SPT) collaboration \citep{Story:2012wx,Hou:2012xq}.
For WMAP7 only and WMAP9 only data, we download the Markov chains from the WMAP website\footnote{WMAP7:http://lambda.gsfc.nasa.gov/product/map/dr4/parameters.cfm}$^,$
\footnote{WMAP9:http://lambda.gsfc.nasa.gov/product/map/dr5/parameters.cfm}.
While using WMAP9+SPT, we obtain the Markov chains by using CosmoMC \citep{Lewis:2002ah} with the data and likelihood code provided 
by WMAP \citep{Bennett:2012fp,Hinshaw:2012fq} and SPT \citep{Story:2012wx,Hou:2012xq,Keisler:2011aw} collaborations.

One can see that the measurements from BOSS CMASS-only dataset are consistent with those from CMB,
and adding CMASS to CMB produces significantly tighter constraints than using CMB data alone.
While adding SPT to WMAP9 in a $\Lambda$CDM model, $\Omega_m$ is decreased as found in \cite{Story:2012wx} (although they used WMAP7). 
It is interesting that the mean values from WMAP9+SPT in a $\Lambda$CDM model are much closer to those from WMAP7 than from WMAP9.

\begin{figure}
\centering
\includegraphics[width=1 \columnwidth,clip,angle=0]{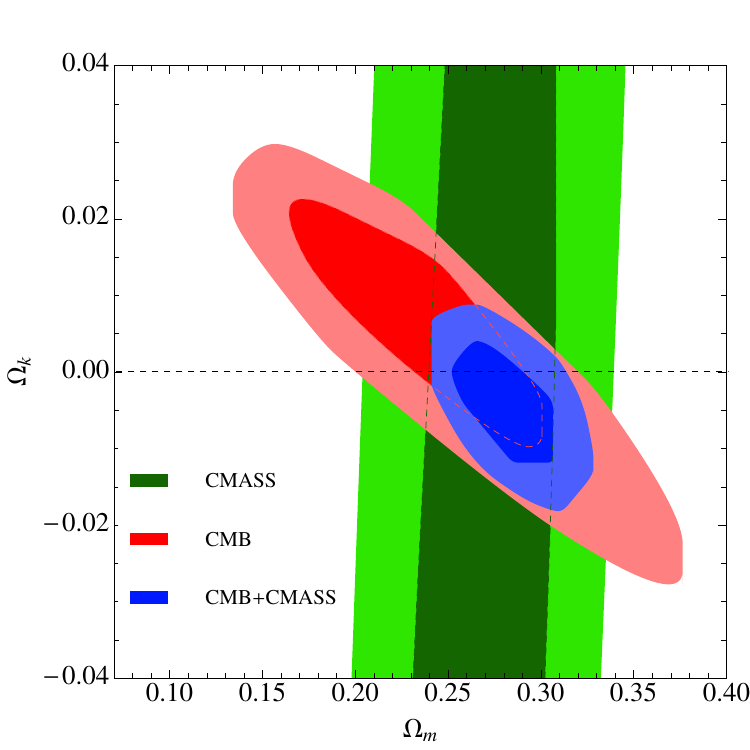}
\caption{
2D marginalized
  contours for $68\%$ and $95\%$ confidence levels for $\Omega_k$ and $\Omega_m$ (o$\Lambda$CDM model assumed)
  from WMAP9+SPT (red), CMASS (green),
  and WMAP9+SPT+CMASS (blue).
The CMASS data break the degeneracy between $\Omega_k$ and $\Omega_m$ constrained by CMB data.
}
\label{fig:om_ok}
\end{figure}

\begin{figure}
\centering
\includegraphics[width=1 \columnwidth,clip,angle=0]{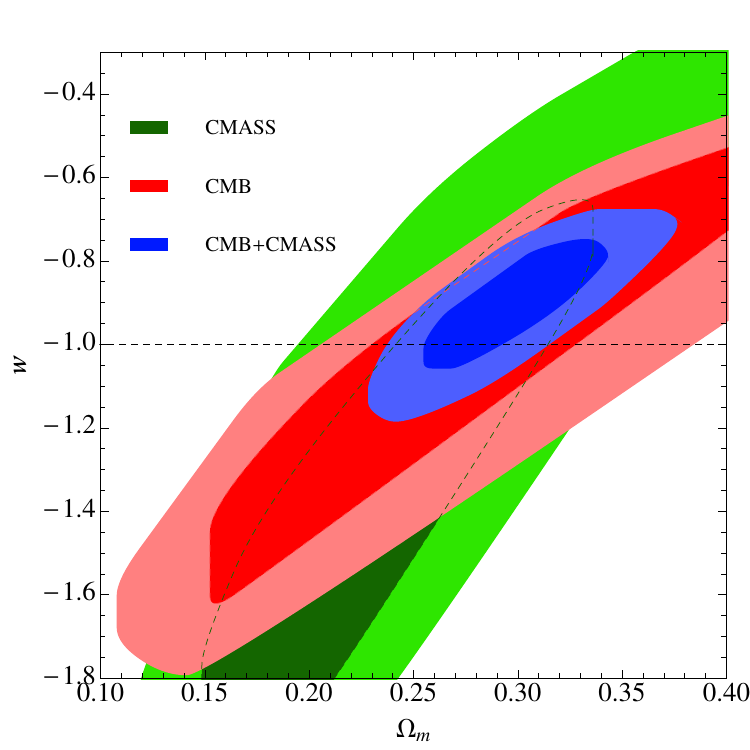}
\caption{
2D marginalized
  contours for $68\%$ and $95\%$ confidence levels for $w$ and $\Omega_m$ ($w$CDM model assumed)
  from WMAP9+SPT (red), CMASS (green),
  and WMAP9+SPT+CMASS (blue).
The CMASS data break the degeneracy between $w$ and $\Omega_m$ constrained by CMB data.
}
\label{fig:om_w}
\end{figure}

\begin{figure}
\centering
\includegraphics[width=1 \columnwidth,clip,angle=0]{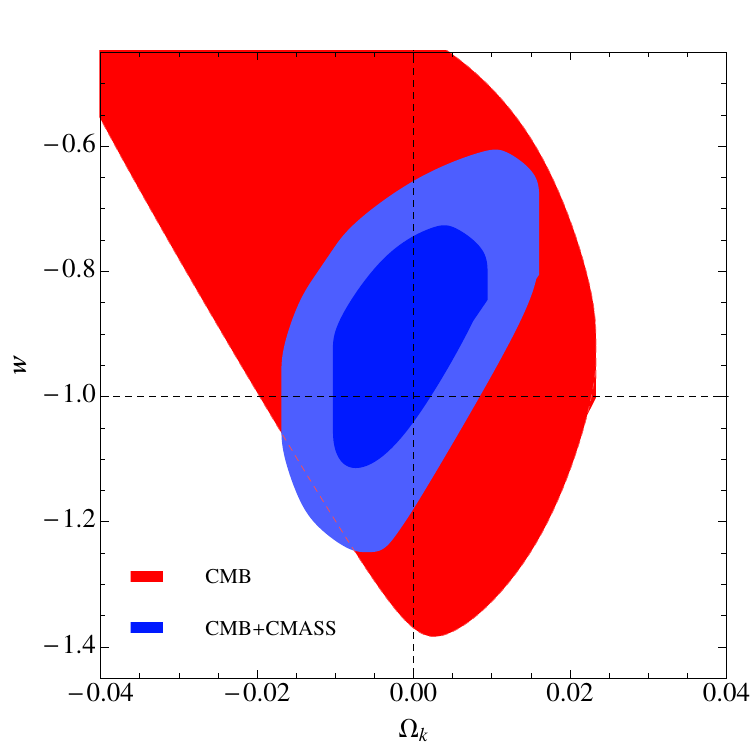}
\caption{
2D marginalized
  contours for $68\%$ and $95\%$ confidence levels for $w$ and $\Omega_k$ (o$w$CDM model assumed)
  from WMAP9+SPT+CMASS (blue). 
  While $w$ and $\Omega_k$ are not well constrained by WMAP9+SPT or CMASS-only, we only show the contour of $68\%$ confidence level from WMAP9+SPT (red).
Adding the CMASS data on the CMB data improves the constraints on $w$ and $\Omega_k$ significantly, and
the results are consistent with $w=-1$ and $\Omega_k=0$ ($\Lambda$CDM model).
}
\label{fig:w_ok}
\end{figure}

Figure \ref{fig:om_ok} shows how CMASS clustering breaks the degeneracy between $\Omega_k$ and $\Omega_m$ constrained by CMB in the o$\Lambda$CDM model, 
resulting in a much tighter constraint.
Figure \ref{fig:om_w} demonstrates how CMASS clustering also breaks the degeneracy between $w$ and $\Omega_m$ constrained by CMB in the $w$CDM model,
resulting in a much better constraint in which $w$ is consistent (within 1 $\sigma$) with $w=-1$ (cosmological constant model). 
This statement is true independent of which
CMB data set (see Table \ref{table:wcdm}).
Figure \ref{fig:w_ok} shows that adding the CMASS and the CMB data improves the constraints on $w$ and $\Omega_k$ significantly in the o$w$CDM model, and 
the results are consistent with $w=-1$ and $\Omega_k=0$ (i.e. a flat $\Lambda$CDM model). This statement holds regardless of which CMB data set is used (see Table \ref{table:owcdm}).

\begin{table*}\scriptsize
\begin{center}
\begin{tabular}{crrrrrrr}
\hline
$\Lambda$CDM		 &CMASS only 		&WMAP7 only 	&WMAP7+CMASS  &WMAP9 only	&WMAP9+CMASS &WMAP9+SPT &WMAP9+SPT+CMASS \\ \hline	
$ \Omega_m $ 	&\ \ $	0.273	\pm	0.032	$	&\ \ $	0.266	\pm	0.029	$	&\ \ $	0.278	\pm	0.019	$	&\ \ $	0.280	\pm	0.026	$	&\ \ $	0.284	\pm	0.017	$	&\ \ $	0.264	\pm	0.019	$	&\ \ $	0.274	\pm	0.015	$	\\
$ H_0 $ 	&\ \ $	68.0	\pm	3.0	$	&\ \ $	71.0	\pm	2.5	$	&\ \ $	69.8	\pm	1.6	$	&\ \ $	70.0	\pm	2.2	$	&\ \ $	69.5	\pm	1.5	$	&\ \ $	71.2	\pm	1.7	$	&\ \ $	70.3	\pm	1.3	$	\\
$ b\sigma_8(0.57) $ 	&\ \ $	1.18	\pm	0.14	$	&\ \ $	-			$	&\ \ $	1.18	\pm	0.11	$	&\ \ $	-			$	&\ \ $	1.19	\pm	0.10	$	&\ \ $	-			$	&\ \ $	1.17	\pm	0.10	$	\\
$ f(0.57)\sigma_8(0.57) $ 	&\ \ $	0.449	\pm	0.055	$	&\ \ $	0.450	\pm	0.025	$	&\ \ $	0.457	\pm	0.018	$	&\ \ $	0.466	\pm	0.019	$	&\ \ $	0.467	\pm	0.015	$	&\ \ $	0.453	\pm	0.016	$	&\ \ $	0.459	\pm	0.012	$	\\
$ b $ 	&\ \ $	2.01	\pm	0.42	$	&\ \ $	-			$	&\ \ $	1.94	\pm	0.18	$	&\ \ $	-			$	&\ \ $	1.93	\pm	0.17	$	&\ \ $	-			$	&\ \ $	1.90	\pm	0.17	$	\\
$ \beta\equiv f(0.57)/b $ 	&\ \ $	0.387	\pm	0.076	$	&\ \ $	-			$	&\ \ $	0.391	\pm	0.035	$	&\ \ $	-			$	&\ \ $	0.396	\pm	0.034	$	&\ \ $	-			$	&\ \ $	0.397	\pm	0.034	$	\\
$ \Omega_mh^2 $ 	&\ \ $	0.125	\pm	0.018	$	&\ \ $	0.1334	\pm	0.0055	$	&\ \ $	0.1351	\pm	0.0039	$	&\ \ $	0.1364	\pm	0.0045	$	&\ \ $	0.1369	\pm	0.0033	$	&\ \ $	0.1336	\pm	0.0035	$	&\ \ $	0.1352	\pm	0.0028	$	\\
$ \sigma_8 $ 	&\ \ $	0.80	\pm	0.10	$	&\ \ $	0.800	\pm	0.030	$	&\ \ $	0.806	\pm	0.024	$	&\ \ $	0.821	\pm	0.023	$	&\ \ $	0.820	\pm	0.020	$	&\ \ $	0.806	\pm	0.018	$	&\ \ $	0.811	\pm	0.015	$	\\
$ f(0.57) $ 	&\ \ $	0.750	\pm	0.030	$	&\ \ $	0.743	\pm	0.025	$	&\ \ $	0.753	\pm	0.016	$	&\ \ $	0.754	\pm	0.021	$	&\ \ $	0.758	\pm	0.014	$	&\ \ $	0.742	\pm	0.017	$	&\ \ $	0.750	\pm	0.013	$	\\
$ H(0.57) $ 	&\ \ $	90.3	\pm	4.6	$	&\ \ $	94.2	\pm	1.3	$	&\ \ $	93.52	\pm	0.95	$	&\ \ $	93.9	\pm	1.1	$	&\ \ $	93.59	\pm	0.83	$	&\ \ $	94.36	\pm	0.80	$	&\ \ $	93.96	\pm	0.61	$	\\
$ D_A(0.57) $ 	&\ \ $	1413	\pm	64	$	&\ \ $	1347	\pm	33	$	&\ \ $	1364	\pm	22	$	&\ \ $	1359	\pm	29	$	&\ \ $	1366	\pm	20	$	&\ \ $	1344	\pm	21	$	&\ \ $	1355	\pm	16	$	\\
$ D_V(0.57) $ 	&\ \ $	2106	\pm	98	$	&\ \ $	2010	\pm	42	$	&\ \ $	2031	\pm	28	$	&\ \ $	2023	\pm	37	$	&\ \ $	2033	\pm	26	$	&\ \ $	2006	\pm	27	$	&\ \ $	2019	\pm	20	$	\\
\hline
\end{tabular}
\end{center}
\caption{
The cosmological constraints using different combinations of data assuming $\Lambda$CDM.
One could see that the measurements from CMASS only are in good agreement with those from CMB.
Combining CMASS with CMB gives significantly tighter constraints than using CMB only.
The unit of $H$ is $\Hunit$. The unit of $D_A$ and $D_V$ is $\rm Mpc$.
} \label{table:lcdm}
\end{table*}

\begin{table*}\scriptsize
\begin{center}
\begin{tabular}{crrrrrrr}
\hline
o$\Lambda$CDM		 &CMASS only 		&WMAP7 only 	&WMAP7+CMASS  &WMAP9 only	&WMAP9+CMASS &WMAP9+SPT &WMAP9+SPT+CMASS \\ \hline
$ \Omega_k $ 	&\ \ $	-0.05	\pm	0.11	$	&\ \ $	-0.083	\pm	0.082	$	&\ \ $	-0.0078	\pm	0.0049	$	&\ \ $	-0.036	\pm	0.059	$	&\ \ $	-0.0048	\pm	0.0067	$	&\ \ $	0.005	\pm	0.012	$	&\ \ $	-0.0043	\pm	0.0054	$	\\
$ \Omega_m $ 	&\ \ $	0.264	\pm	0.033	$	&\ \ $	0.58	\pm	0.30	$	&\ \ $	0.288	\pm	0.019	$	&\ \ $	0.43	\pm	0.23	$	&\ \ $	0.289	\pm	0.022	$	&\ \ $	0.242	\pm	0.052	$	&\ \ $	0.282	\pm	0.019	$	\\
$H_0 $ 	&\ \ $	68.8	\pm	4.0	$	&\ \ $	53	\pm	13	$	&\ \ $	67.5	\pm	1.9	$	&\ \ $	62	\pm	14	$	&\ \ $	68.5	\pm	2.3	$	&\ \ $	75.7	\pm	7.9	$	&\ \ $	68.9	\pm	2.2	$	\\
$ b\sigma_8(0.57) $ 	&\ \ $	1.19	\pm	0.14	$	&\ \ $	-			$	&\ \ $	1.23	\pm	0.11	$	&\ \ $	-			$	&\ \ $	1.19	\pm	0.12	$	&\ \ $	-			$	&\ \ $	1.19	\pm	0.11	$	\\
$ f(0.57)\sigma_8(0.57) $ 	&\ \ $	0.440	\pm	0.058	$	&\ \ $	0.467	\pm	0.028	$	&\ \ $	0.451	\pm	0.020	$	&\ \ $	0.475	\pm	0.029	$	&\ \ $	0.466	\pm	0.018	$	&\ \ $	0.442	\pm	0.024	$	&\ \ $	0.458	\pm	0.012	$	\\
$ b $ 	&\ \ $	2.10	\pm	0.49	$	&\ \ $	-			$	&\ \ $	2.09	\pm	0.21	$	&\ \ $	-			$	&\ \ $	1.96	\pm	0.23	$	&\ \ $	-			$	&\ \ $	1.98	\pm	0.19	$	\\
$ \beta\equiv f(0.57)/b $ 	&\ \ $	0.378	\pm	0.076	$	&\ \ $	-			$	&\ \ $	0.369	\pm	0.036	$	&\ \ $	-			$	&\ \ $	0.395	\pm	0.042	$	&\ \ $	-			$	&\ \ $	0.386	\pm	0.035	$	\\
$ \Omega_mh^2 $ 	&\ \ $	0.125	\pm	0.019	$	&\ \ $	0.1336	\pm	0.0056	$	&\ \ $	0.1308	\pm	0.0050	$	&\ \ $	0.1371	\pm	0.0045	$	&\ \ $	0.1350	\pm	0.0048	$	&\ \ $	0.1340	\pm	0.0036	$	&\ \ $	0.1338	\pm	0.0034	$	\\
$ \sigma_8 $ 	&\ \ $	0.77	\pm	0.11	$	&\ \ $	0.760	\pm	0.046	$	&\ \ $	0.787	\pm	0.030	$	&\ \ $	0.805	\pm	0.036	$	&\ \ $	0.814	\pm	0.026	$	&\ \ $	0.808	\pm	0.020	$	&\ \ $	0.803	\pm	0.018	$	\\
$ f(0.57) $ 	&\ \ $	0.760	\pm	0.040	$	&\ \ $	0.90	\pm	0.12	$	&\ \ $	0.764	\pm	0.016	$	&\ \ $	0.83	\pm	0.11	$	&\ \ $	0.764	\pm	0.018	$	&\ \ $	0.714	\pm	0.053	$	&\ \ $	0.758	\pm	0.016	$	\\
$ H(0.57) $ 	&\ \ $	89.0	\pm	5.0	$	&\ \ $	80.7	\pm	9.8	$	&\ \ $	90.9	\pm	1.8	$	&\ \ $	88	\pm	10	$	&\ \ $	92.4	\pm	2.1	$	&\ \ $	98.3	\pm	6.6	$	&\ \ $	92.5	\pm	1.9	$	\\
$ D_A(0.57) $ 	&\ \ $	1408	\pm	67	$	&\ \ $	1735	\pm	309	$	&\ \ $	1407	\pm	31	$	&\ \ $	1535	\pm	268	$	&\ \ $	1385	\pm	38	$	&\ \ $	1286	\pm	110	$	&\ \ $	1379	\pm	36	$	\\
$ D_V(0.57) $ 	&\ \ $	2112	\pm	101	$	&\ \ $	2516	\pm	397	$	&\ \ $	2094	\pm	44	$	&\ \ $	2252	\pm	349	$	&\ \ $	2061	\pm	53	$	&\ \ $	1924	\pm	153	$	&\ \ $	2054	\pm	50	$	\\
\hline
\end{tabular}
\end{center}
\caption{Same as Table \ref{table:lcdm} but
assuming o$\Lambda$CDM. 
The constraints from CMASS only are obtained with the flat prior (0.55,1) on the density fraction of dark energy, $\Omega_w(=1-\Omega_m-\Omega_k)$ .
We note that the constraints on the curvature, $\Omega_k$, is  in a good agreement with a flat universe ($\Omega_k=0$).
} \label{table:olcdm}
\end{table*}

\begin{table*}\scriptsize
\begin{center}
\begin{tabular}{crrrrrrr}
\hline
$w$CDM		 &CMASS only 		&WMAP7 only 	&WMAP7+CMASS  &WMAP9 only	&WMAP9+CMASS &WMAP9+SPT &WMAP9+SPT+CMASS \\ \hline	
$ w $ 	&\ \ $	-1.24	\pm	0.42	$	&\ \ $	-1.09	\pm	0.38	$	&\ \ $	-0.94	\pm	0.12	$	&\ \ $	-1.01	\pm	0.43	$	&\ \ $	-0.94	\pm	0.13	$	&\ \ $	-0.97	\pm	0.37	$	&\ \ $	-0.90	\pm	0.11	$	\\
$ \Omega_m $ 	&\ \ $	0.247	\pm	0.064	$	&\ \ $	0.264	\pm	0.098	$	&\ \ $	0.290	\pm	0.029	$	&\ \ $	0.30	\pm	0.11	$	&\ \ $	0.295	\pm	0.034	$	&\ \ $	0.30	\pm	0.11	$	&\ \ $	0.297	\pm	0.031	$	\\
$ H_0 $ 	&\ \ $	73.0	\pm	9.2	$	&\ \ $	75	\pm	13	$	&\ \ $	68.3	\pm	3.4	$	&\ \ $	71	\pm	14	$	&\ \ $	68.1	\pm	3.8	$	&\ \ $	71	\pm	13	$	&\ \ $	67.5	\pm	3.5	$	\\
$ b\sigma_8(0.57) $ 	&\ \ $	1.20	\pm	0.14	$	&\ \ $	-			$	&\ \ $	1.20	\pm	0.11	$	&\ \ $	-			$	&\ \ $	1.19	\pm	0.10	$	&\ \ $	-			$	&\ \ $	1.21	\pm	0.11	$	\\
$ f(0.57)\sigma_8(0.57) $ 	&\ \ $	0.435	\pm	0.065	$	&\ \ $	0.471	\pm	0.091	$	&\ \ $	0.442	\pm	0.033	$	&\ \ $	0.47	\pm	0.10	$	&\ \ $	0.450	\pm	0.035	$	&\ \ $	0.450	\pm	0.079	$	&\ \ $	0.436	\pm	0.029	$	\\
$ b $ 	&\ \ $	2.20	\pm	0.58	$	&\ \ $	-			$	&\ \ $	2.04	\pm	0.26	$	&\ \ $	-			$	&\ \ $	2.00	\pm	0.25	$	&\ \ $	-			$	&\ \ $	2.08	\pm	0.26	$	\\
$ \beta\equiv f(0.57)/b $ 	&\ \ $	0.370	\pm	0.082	$	&\ \ $	-			$	&\ \ $	0.372	\pm	0.049	$	&\ \ $	-			$	&\ \ $	0.381	\pm	0.050	$	&\ \ $	-			$	&\ \ $	0.364	\pm	0.048	$	\\
$ \Omega_mh^2 $ 	&\ \ $	0.127	\pm	0.019	$	&\ \ $	0.1335	\pm	0.0056	$	&\ \ $	0.1342	\pm	0.0039	$	&\ \ $	0.1364	\pm	0.0048	$	&\ \ $	0.1357	\pm	0.0037	$	&\ \ $	0.1339	\pm	0.0036	$	&\ \ $	0.1343	\pm	0.0030	$	\\
$ \sigma_8 $ 	&\ \ $	0.75	\pm	0.15	$	&\ \ $	0.82	\pm	0.13	$	&\ \ $	0.785	\pm	0.046	$	&\ \ $	0.82	\pm	0.15	$	&\ \ $	0.796	\pm	0.048	$	&\ \ $	0.79	\pm	0.12	$	&\ \ $	0.778	\pm	0.041	$	\\
$ f(0.57) $ 	&\ \ $	0.773	\pm	0.055	$	&\ \ $	0.752	\pm	0.029	$	&\ \ $	0.747	\pm	0.016	$	&\ \ $	0.764	\pm	0.027	$	&\ \ $	0.752	\pm	0.017	$	&\ \ $	0.750	\pm	0.020	$	&\ \ $	0.745	\pm	0.014	$	\\
$ H(0.57) $ 	&\ \ $	88.4	\pm	6.1	$	&\ \ $	93.1	\pm	1.9	$	&\ \ $	93.8	\pm	1.0	$	&\ \ $	92.6	\pm	1.9	$	&\ \ $	93.72	\pm	0.91	$	&\ \ $	93.3	\pm	1.4	$	&\ \ $	93.99	\pm	0.61	$	\\
$ D_A(0.57) $ 	&\ \ $	1400	\pm	69	$	&\ \ $	1346	\pm	104	$	&\ \ $	1375	\pm	30	$	&\ \ $	1384	\pm	115	$	&\ \ $	1377	\pm	33	$	&\ \ $	1374	\pm	110	$	&\ \ $	1379	\pm	31	$	\\
$ D_V(0.57) $ 	&\ \ $	2109	\pm	101	$	&\ \ $	2015	\pm	108	$	&\ \ $	2040	\pm	33	$	&\ \ $	2057	\pm	119	$	&\ \ $	2043	\pm	35	$	&\ \ $	2042	\pm	114	$	&\ \ $	2043	\pm	33	$	\\
\hline
\end{tabular}
\end{center}
\caption{Same as Table \ref{table:lcdm} but
assuming $w$CDM. 
The constraints from CMASS only are obtained with the flat priors (-2,0) on the constant of equation of state of dark energy, $w$.
One can see that the constraints on $w$ is consistent with -1 (cosmological constant model).
} \label{table:wcdm}
\end{table*}

\begin{table*}\scriptsize
\begin{center}
\begin{tabular}{crrrrrrr}
\hline
o$w$CDM		 &CMASS only 		&WMAP7 only 	&WMAP7+CMASS  &WMAP9 only	&WMAP9+CMASS &WMAP9+SPT &WMAP9+SPT+CMASS \\ \hline	
$ w $ 	&\ \ $	-1.19	\pm	0.45	$	&\ \ $	-1.18	\pm	0.60	$	&\ \ $	-0.99	\pm	0.21	$	&\ \ $	-0.88	\pm	0.59	$	&\ \ $	-0.93	\pm	0.14	$	&\ \ $	-0.77	\pm	0.35	$	&\ \ $	-0.91	\pm	0.13	$	\\
$ \Omega_k $ 	&\ \ $	-0.02	\pm	0.13	$	&\ \ $	-0.105	\pm	0.084	$	&\ \ $	-0.0017	\pm	0.0084	$	&\ \ $	-0.053	\pm	0.063	$	&\ \ $	-0.0010	\pm	0.0067	$	&\ \ $	-0.009	\pm	0.025	$	&\ \ $	-0.0014	\pm	0.0071	$	\\
$ \Omega_m $ 	&\ \ $	0.258	\pm	0.061	$	&\ \ $	0.67	\pm	0.32	$	&\ \ $	0.297	\pm	0.044	$	&\ \ $	0.56	\pm	0.25	$	&\ \ $	0.302	\pm	0.033	$	&\ \ $	0.38	\pm	0.18	$	&\ \ $	0.299	\pm	0.030	$	\\
$ H_0 $ 	&\ \ $	70.9	\pm	8.3	$	&\ \ $	50	\pm	16	$	&\ \ $	67.9	\pm	4.8	$	&\ \ $	54	\pm	14	$	&\ \ $	67.4	\pm	3.7	$	&\ \ $	65	\pm	16	$	&\ \ $	67.1	\pm	3.3	$	\\
$ b\sigma_8(0.57) $ 	&\ \ $	1.19	\pm	0.14	$	&\ \ $	-			$	&\ \ $	1.27	\pm	0.11	$	&\ \ $	-			$	&\ \ $	1.21	\pm	0.11	$	&\ \ $	-			$	&\ \ $	1.22	\pm	0.11	$	\\
$ f(0.57)\sigma_8(0.57) $ 	&\ \ $	0.442	\pm	0.063	$	&\ \ $	0.484	\pm	0.077	$	&\ \ $	0.454	\pm	0.042	$	&\ \ $	0.45	\pm	0.10	$	&\ \ $	0.450	\pm	0.039	$	&\ \ $	0.412	\pm	0.066	$	&\ \ $	0.436	\pm	0.032	$	\\
$ b $ 	&\ \ $	2.12	\pm	0.57	$	&\ \ $	-			$	&\ \ $	2.16	\pm	0.28	$	&\ \ $	-			$	&\ \ $	2.04	\pm	0.26	$	&\ \ $	-			$	&\ \ $	2.11	\pm	0.25	$	\\
$ \beta\equiv f(0.57)/b $ 	&\ \ $	0.380	\pm	0.081	$	&\ \ $	-			$	&\ \ $	0.361	\pm	0.058	$	&\ \ $	-			$	&\ \ $	0.376	\pm	0.049	$	&\ \ $	-			$	&\ \ $	0.360	\pm	0.047	$	\\
$ \Omega_mh^2 $ 	&\ \ $	0.126	\pm	0.019	$	&\ \ $	0.1345	\pm	0.0056	$	&\ \ $	0.1346	\pm	0.0048	$	&\ \ $	0.1365	\pm	0.0048	$	&\ \ $	0.1360	\pm	0.0039	$	&\ \ $	0.1339	\pm	0.0037	$	&\ \ $	0.1338	\pm	0.0035	$	\\
$ \sigma_8 $ 	&\ \ $	0.78	\pm	0.14	$	&\ \ $	0.76	\pm	0.11	$	&\ \ $	0.793	\pm	0.051	$	&\ \ $	0.74	\pm	0.14	$	&\ \ $	0.794	\pm	0.050	$	&\ \ $	0.73	\pm	0.11	$	&\ \ $	0.775	\pm	0.042	$	\\
$ f(0.57) $ 	&\ \ $	0.762	\pm	0.056	$	&\ \ $	0.94	\pm	0.11	$	&\ \ $	0.764	\pm	0.027	$	&\ \ $	0.87	\pm	0.10	$	&\ \ $	0.756	\pm	0.025	$	&\ \ $	0.760	\pm	0.064	$	&\ \ $	0.748	\pm	0.021	$	\\
$ H(0.57) $ 	&\ \ $	89.2	\pm	6.0	$	&\ \ $	-			$	&\ \ $	92.1	\pm	3.6	$	&\ \ $	-			$	&\ \ $	93.4	\pm	2.5	$	&\ \ $	-			$	&\ \ $	93.5	\pm	2.4	$	\\
$ D_A(0.57) $ 	&\ \ $	1406	\pm	69	$	&\ \ $	-			$	&\ \ $	1396	\pm	41	$	&\ \ $	-			$	&\ \ $	1387	\pm	41	$	&\ \ $	-			$	&\ \ $	1388	\pm	35	$	\\
$ D_V(0.57) $ 	&\ \ $	2108	\pm	101	$	&\ \ $	-			$	&\ \ $	2075	\pm	58	$	&\ \ $	-			$	&\ \ $	2055	\pm	54	$	&\ \ $	-			$	&\ \ $	2056	\pm	47	$	\\
\hline
\end{tabular}
\end{center}
\caption{Same as Table \ref{table:lcdm} but
assuming o$w$CDM. 
The constraints from CMASS only are obtained with the flat priors (0.55,1) and (-2,0) on $\Omega_w$ and $w$ respectively.
In all the cases, the constraints are consistent with $w=-1$ and $\Omega_k=0$ ($\Lambda$CDM model).
} \label{table:owcdm}
\end{table*}

\subsection{Power of the Constraints on dark energy from $f(z)\sigma_8(z)$} 
\label{sec:fsigma8}
In this section, we demonstrate how adding the measurement of $f(z)\sigma_8(z)$ from galaxy clustering data would improve the constraints on the cosmological parameters.
\cite{Samushia:2012iq}, using the BOSS CMASS DR9 measurements from \cite{Reid:2012sw}, found that the extra information from the 2D correlation function 
compared to the spherically-averaged correlation function improves the constraint on $w$ significantly in the $w$CDM model.
The extra information from the anisotropic galaxy clustering includes 
the geometric distortion, also called Alcock-Paczynski effect \citep{Alcock:1979mp}, and the redshift space distortion (RSD). 
In Table. \ref{table:diff_cmass}, we combine CMB (WMAP9+SPT) with different portions of the information obtained from our BOSS CMASS galaxy clustering analysis.
First, we use $D_V(z)/r_s$, which is the main measurement from the spherically-averaged correlation function. 
Second, we use $H(z)r_s$+$D_A(z)/r_s$, which could be considered as adding geometric distortion on the previous one.
Third, we use $H(z)r_s$+$D_A(z)/r_s$+$f(z)\sigma_8(z)$, which consists of adding the RSD information.
We find that adding $f(z)\sigma_8(z)$ actually dominates the improvement of the cosmological parameter constraints;
but there is only a small difference between using $H(z)r_s$+$D_A(z)/r_s$ and using $D_V(z)/r_s$.
There is no significant improvement between using $H(z)r_s$+$D_A(z)/r_s$ and using $D_V(z)/r_s$ measured from SDSS DR7 LRGs \citep{Wang:2011sb,Xu:2012fw}.
\cite{Anderson13} also find the similar results from BOSS DR9 CMASS galaxy sample.

\begin{figure}
\centering
\includegraphics[width=1 \columnwidth,clip,angle=0]{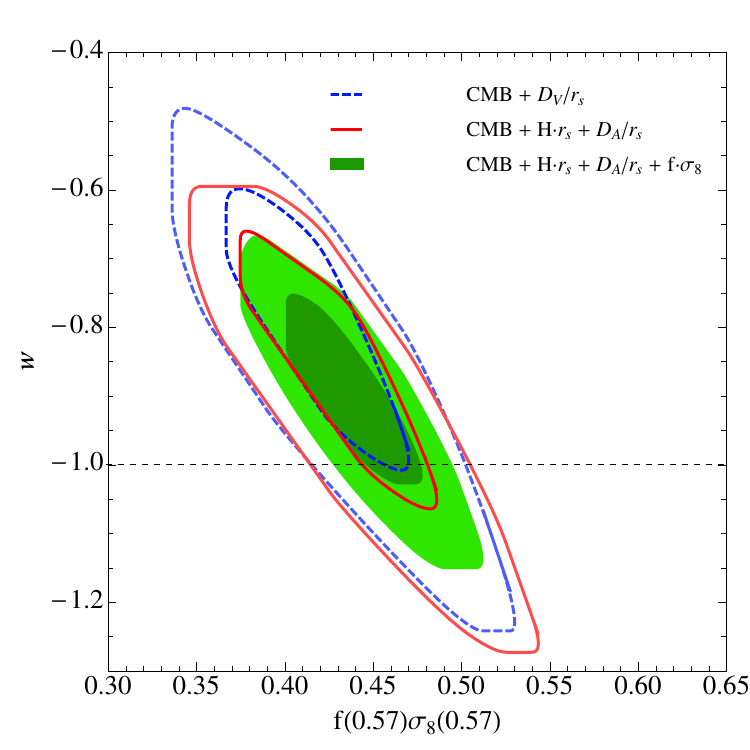}
\caption{
2D marginalized
  contours for $68\%$ and $95\%$ confidence levels for $w$ and $f(z)\sigma_8(z)$ ($w$CDM model assumed)
  from CMB+$D_V(z)/r_s$ (thin solid red), CMB+$H(z) r_s$+$D_A(z)/r_s$CMASS(dashed blue).
  and CMB+$H(z) r_s$+$D_A(z)/r_s$+$f(z)\sigma_8(z)$ (thick solid black).
One can see that there is no much improvement while replacing $D_V(z)/r_s$ with $H(z)r_s$+$D_A(z)/r_s$, 
but have significantly improvement on dark energy while adding $f(z)\sigma_8(z)$.  
}
\label{fig:fs8_w_diff_cmass}
\end{figure}

Figure \ref{fig:fs8_w_diff_cmass} shows the constraints obtained from using different portions of the information from galaxy clustering as described above. 
There is not much difference when replacing $D_V(z)/r_s$ with $H(z)r_s$ and $D_A(z)/r_s$, but significant improvement when adding $f(z)\sigma_8(z)$.
As explained in \cite{Samushia:2012iq}, the correlation between $f(z)\sigma_8(z)$ and the geometric distortion (Alcock-Paczynski effect) 
contributes to the improvement of the constraint on dark energy as well.

\begin{table*}\scriptsize
\begin{center}
\begin{tabular}{crrr|rrr}
\hline
\toprule
&\multicolumn{3}{c|}{$w$CDM} &\multicolumn{3}{c}{o$w$CDM}\\
\midrule
		 &CMB+$D_V/r_s$ 		&CMB+$H r_s$+$D_A/r_s$	  &CMB+$H r_s$+$D_A/r_s$+$f\sigma_8$	&CMB+$D_V/r_s$ 		&CMB+$H r_s$+$D_A/r_s$	  &CMB+$H r_s$+$D_A/r_s$+$f\sigma_8$ \\ \hline	
$ w $ 	&\ \ $	-0.83	\pm	0.17	$	&\ \ $	-0.89	\pm	0.16	$	&\ \ $	-0.90	\pm	0.10	$	&\ \ $	-0.85	\pm	0.26	$	&\ \ $	-0.93	\pm	0.26	$	&\ \ $	-0.93	\pm	0.14	$	\\
$\Omega_k$	&\ \ $	-			$	&\ \ $	-			$	&\ \ $	-			$	&\ \ $	0.0007	\pm	0.0088	$	&\ \ $	-0.0007	\pm	0.0081	$	&\ \ $	-0.0019	\pm	0.0066	$	\\
$ \Omega_m $ 	&\ \ $	0.324	\pm	0.043	$	&\ \ $	0.306	\pm	0.036	$	&\ \ $	0.300	\pm	0.027	$	&\ \ $	0.322	\pm	0.053	$	&\ \ $	0.301	\pm	0.047	$	&\ \ $	0.299	\pm	0.029	$	\\
$ H_0 $ 	&\ \ $	64.9	\pm	4.9	$	&\ \ $	66.7	\pm	4.5	$	&\ \ $	67.2	\pm	3.1	$	&\ \ $	65.2	\pm	6.2	$	&\ \ $	67.4	\pm	5.9	$	&\ \ $	67.2	\pm	3.2	$	\\
$ f(0.57)\sigma_8(0.57) $ 	&\ \ $	0.420	\pm	0.044	$	&\ \ $	0.433	\pm	0.043	$	&\ \ $	0.438	\pm	0.029	$	&\ \ $	0.423	\pm	0.064	$	&\ \ $	0.442	\pm	0.064	$	&\ \ $	0.442	\pm	0.035	$	\\
$ \Omega_mh^2 $ 	&\ \ $	0.1342	\pm	0.0035	$	&\ \ $	0.1345	\pm	0.0036	$	&\ \ $	0.1349	\pm	0.0031	$	&\ \ $	0.1339	\pm	0.0036	$	&\ \ $	0.1340	\pm	0.0036	$	&\ \ $	0.1342	\pm	0.0036	$	\\
$ \sigma_8 $ 	&\ \ $	0.751	\pm	0.063	$	&\ \ $	0.772	\pm	0.060	$	&\ \ $	0.780	\pm	0.041	$	&\ \ $	0.752	\pm	0.084	$	&\ \ $	0.779	\pm	0.081	$	&\ \ $	0.782	\pm	0.045	$	\\
$ f(0.57) $ 	&\ \ $	0.746	\pm	0.016	$	&\ \ $	0.747	\pm	0.017	$	&\ \ $	0.748	\pm	0.014	$	&\ \ $	0.749	\pm	0.029	$	&\ \ $	0.753	\pm	0.031	$	&\ \ $	0.753	\pm	0.021	$	\\
$ H(0.57) $ 	&\ \ $	93.68	\pm	0.70	$	&\ \ $	93.81	\pm	0.70	$	&\ \ $	93.91	\pm	0.61	$	&\ \ $	93.3	\pm	2.8	$	&\ \ $	93.0	\pm	3.0	$	&\ \ $	93.1	\pm	2.3	$	\\
$ D_A(0.57) $ 	&\ \ $	1407	\pm	44	$	&\ \ $	1389	\pm	37	$	&\ \ $	1382	\pm	28	$	&\ \ $	1410	\pm	43	$	&\ \ $	1392	\pm	36	$	&\ \ $	1390	\pm	30	$	\\
$ D_V(0.57) $ 	&\ \ $	2073	\pm	45	$	&\ \ $	2054	\pm	38	$	&\ \ $	2047	\pm	29	$	&\ \ $	2078	\pm	46	$	&\ \ $	2064	\pm	42	$	&\ \ $	2060	\pm	40	$	\\
\hline
\end{tabular}
\end{center}
\caption{
The cosmological constraints using CMB (WMAP9+SPT) and different portions of the information extracted from CMASS assuming $w$CDM and o$w$CDM.
The unit of $H$ is $\Hunit$ and the unit of $D_A$ and $D_V$ is $\rm Mpc$.
} \label{table:diff_cmass}
\end{table*}

\section{COMPARISON WITH PREVIOUS WORKS}
\label{sec:compare}

We compare our results with previous or parallel works which use the same data (BOSS CMASS DR9). In general, although the data set used are identical, the results 
could be slightly different due to using different methodology, models, or parameter spaces explored.

\cite{Anderson:2012sa} measured $\alpha=1.016\pm0.017$ from the position of the BAO peak of the two-point correlation function (monopole), before applying 
the reconstruction method on the density field. 
\cite{Sanchez:2012sg} measured $\alpha=1.015\pm0.019$ from the full shape of the correlation function (monopole). 
\cite{Reid:2012sw} obtain $\alpha=1.023\pm0.019$ from the broad-range shape of the monopole and quadrupole of the correlation function; 
and \cite{Ross:2012qm} obtain $\alpha=1.020\pm0.019$ using monopole and quadrupole as well. These results are all consistent with our result $\alpha=1.024\pm0.022$.
Our uncertainty is slightly larger because we marginalize over wide ranges of $\Omega_b h^2$ and $n_s$ as mentioned in Sec \ref{sec:mcmc}.

Comparing to our parallel works, \cite{Anderson13} 
obtain $D_A(0.57)/r_s = 9.20\pm0.29$ and $H(0.57)r_s/c=0.0474\pm0.0040$;
\cite{Kazin13} obtain $D_A(0.57)/r_s = 9.05\pm0.27$ and $H(0.57)r_s/c=0.0464\pm0.0030$; 
both are in excellent agreement with our measurements, $D_A(0.57)/r_s = 8.95\pm0.27$ and $H(0.57)r_s/c=0.0454\pm0.0031$, 
in spite of the differences among different methodology, models, and analysis.
\cite{Sanchez13}, our other parallel work, combines CMASS and other datasets (CMB, SNe, etc) to measure cosmological parameters adopting specific dark energy models. 
Here, we only discuss the cases of combining CMASS and CMB datasets for comparison.
When combining CMASS and CMB for $\Lambda$CDM, they obtain $\Omega_m=0.285\pm0.015$ and $H_0=69.5\pm1.2$, which agree to our results $\Omega_m=0.274\pm0.015$ and $H_0=70.3\pm1.3$. In the case of a more general model, i.e. o$w$CDM, they obtain $\Omega_k=-0.0023\pm0.0061$ and $w=-0.97\pm0.16$, which results are consistent with ours $\Omega_k=-0.0014\pm0.0071$ and $w=-0.91\pm0.13$. Note, that in spite of 
different model, analysis, and CMB dataset, their results are in good agreement with ours.

\cite{Reid:2012sw} obtained $\{f(0.57)\sigma_8(0.57)$, $F(0.57)$, $D_V(z0.57)/r_s\}$ =$\{0.4298\pm0.0672$, $0.6771\pm0.0417$, $1.0227\pm0.0188\}$ where
$F(z)\equiv (1+z)D_A(z)H(z)/c.$
Our corresponding measurements are $\{0.428\pm0.069$, $0.637\pm0.057$, $1.024\pm0.022\}$. $F(z)$, in our analysis, can be derived from $\epsilon$, see Eq. (\ref{eq:epsilon}). 
Although the results are in good agreement, their 6.2\% error on $F(z)$ is significantly smaller than ours (8.9\%). 
The reason should be that they included smaller scales in their analysis, thus they could obtain tighter constraints. While the uncertainties from galaxy bias, redshift space distortion, and nonlinear effect have larger effects on small scales, we use the scale range which is not sensitive to these uncertainties to obtain robust cosmological measurements from the galaxy sample.

Regarding other results, \cite{Blake:2012pj} measured $\{f(z)\sigma_8(z)$, $H(z)$, $D_A(z)\}$ = $\{0.390\pm0.063$, $87.9\pm6.1$, $1380\pm95\}$ at $z=0.6$ 
using the galaxy sample from WiggleZ \citep{Drinkwater:2009sd}, which is consistent with our measurements $\{0.428\pm0.069$, $87.6\pm7.2$, $1396\pm74\}$ at $z=0.57$.

\section{SINGLE-PROBE MEASUREMENTS: REQUIREMENTS}
\label{sec:single-probe}

In general, an ideal single-probe cosmological measurement should not adopt any cosmological parameter priors from another data set (e.g., CMB, SNe, ...). 
Otherwise, one should not be able to combine the results with other cosmological probe data set to avoid double counting. 
However, in practice, some priors are needed because not all the cosmological parameters could be well constrained by the given observed data.
In other words, the data set can not constrain all the parameters of a given model, so that we need to use priors on some of them.
In addition, adopting priors can reduce computing time significantly, which might be important when using a complicate theoretical model.
In this section, we discuss how priors may be chosen in order to avoid biases in the measurements and underestimations of the uncertainties.

Which cosmological parameters are appropriate to be set on the priors? The first choice would be the parameters which are tightly constrained by other data sets. 
For example, the uncertainties of $n_s$ and $\Omega_b h^2$ measured by CMB could be lower than $1.5\%$ and $2.5\%$ respectively. 
Therefore, the priors on these two parameters are often used in the analyses of galaxy clustering data sets.

What prior we should use for a given parameter? The key is that we have to make sure that the prior will not result in double counting when later we combine with other probes.
This can be done by using a sufficiently wide flat prior. For example, in this study we use $\pm7\sigma_{WMAP}$ for $n_s$ and $\Omega_b h^2$; where $\sigma_{WMAP}$ is the uncertainty measured from WMAP7.

Due to some practical concerns (e.g., the data set size or computing time), one might want to use tighter priors or to adopting priors for more parameters. 
It is fine if the quantities one is measuring are insensitive to those parameters with priors. 
For example, as shown in Table \ref{table:covar_matrix_fid}, the correlation between $D_V(z)/r_s(z_d)$ and $\Omega_m h^2$ is close to zero (i.e. -0.0200), 
so that one could put a strong prior on $\Omega_m h^2$ (i.e., CMB prior) and measure $D_V(z)/r_s(z_d)$ without introducing a bias or underestimation. 
Yet, the proper quantities to be measured might vary with different analysis conditions. For example, \cite{Blake:2011wn} found that $A_s(z)$, instead of $D_V(z)/r_s(z_d)$, is uncorrelated to $\Omega_m h^2$ while including smaller scales (see Fig. 6 and 15 in their paper). 
In other words, when including small scales, $D_V(z)/r_s(z_d)$ is no longer a good measurement while using $\Omega_m h^2$ prior. 
It is worth to mention that \cite{Blake:2011wn} marginalized over $\Omega_m h^2$, which is similar as we did in our study, 
so that both $A_s(z)$ and $D_V(z)/r_s(z_d)$ are fine measurements in their study (they recommended to use $A_s(z)$ as their default result).

In this study, we provide the single-probe clustering measurements by using sufficient wide priors on the parameters, which are not well constrained by BOSS CMASS data.
In addition, it is worth to emphasize that our results are obtained without assuming a dark energy model. Thus, one can derive cosmological parameter constraints of a specific dark energy model from our results or combining our results with other datasets (see examples in Sec. \ref{sec:models}).

\section{SYSTEMATIC TESTS} 
\label{sec:test}

\cite{Ross:2012qm} studied the systematics of the BOSS CMASS data selection, observation, selection functions of the geometry of the survey. 
In this section, we focus on the possible systematics from the theoretical model due to the scale-dependent uncertainties including nonlinear effects, 
galaxy bias and redshift distortions. These effects mainly affect the small scales and they might actually be cancelled by each other. 
For example, the galaxy bias tends to be larger at smaller scales \citep{Nuza:2012mw}, 
but the nonlinear effect makes the correlation function smaller at the same scale (e.g. see Fig 3 in \citealt{Chuang:2010dv}). 
Therefore, if one only corrects the model with one of the effects (i.e. nonlinear effect), the corrected model could become less accurate. 
Here, instead of adding corrections, we use the scale range which is not sensitive to these uncertainties to obtain robust cosmological measurements from the galaxy sample.
Table\ \ref{table:test} shows the systematic tests that we have
performed by varying the scale range used. We vary the lower limit from 25 to 50 $h^{-1}$Mpc and the upper limit from 120 to 200 $h^{-1}$Mpc. 
We find that the results are insensitive to those scale limits, which demonstrates that our constraints are robust to the combination of scale-dependent uncertainties mentioned above.

We also apply our method on the BOSS CMASS sample in the north Galactic hemispheres and south Galactic hemispheres separately. We find that $\alpha=0.996^{+0.028}_{-0.025}$ and $1.085^{+0.030}_{-0.027}$ for 
north and south, which is in excellent agreement with what \cite{Ross:2012qm} obtained, i.e., $\alpha=0.994\pm0.023$ and $1.083\pm0.029$. 
While the deviations between the results from north and south samples are large, \cite{Ross:2012qm} found that it is probably due to the statistical variance instead of 
the presence of some systematic error.

Last, we test the stability of the covariance matrix used by using half number of the mock catalogs (600/2=300) and find the results are insensitive to it. 
The $\chi^2$/d.o.f is 0.57 which is similar to our fiducial result.

\begin{table*}
\begin{center}
\begin{tabular}{crrrrrr}
\hline
&$H$ &$D_A$ &$\Omega_m h^2$ &$\beta$ &$b\sigma_8$ &$r_s H/c$ \\ \hline

fiducial result	
&\ \ $	 87.6	_{	-6.8	}^{+	6.7	}$
&\ \ $	 1396	\pm73$
&\ \ $	 0.126	_{	-0.010	}^{+	0.008	}$
&\ \ $	 0.367	_{	-0.081	}^{+	0.079	}$
&\ \ $	 1.19	\pm0.14$
&\ \ $	 0.0454\pm0.0028$
\\

$25<s<160$
&\ \ $	 88.3	_{	-6.4	}^{+	6.5	}$
&\ \ $	 1386	_{	-66	}^{+	65	}$
&\ \ $	 0.1293	_{	-0.0061	}^{+	0.0055	}$
&\ \ $	 0.353	_{	-0.063	}^{+	0.062	}$
&\ \ $	 1.198	_{	-0.055	}^{+	0.089	}$
&\ \ $	 0.0455	\pm0.0029$
\\

$50<s<160$
&\ \ $	 88.3	_{	-7.1	}^{+	7.0	}$
&\ \ $	 1398	_{	-83	}^{+	81	}$
&\ \ $	 0.128	_{	-0.013	}^{+	0.009	}$
&\ \ $	 0.37	\pm0.10$
&\ \ $	 1.19	\pm0.21$
&\ \ $	 0.0457	_{	-0.0028	}^{+	0.0027	}$
\\

$40<s<120$
&\ \ $	88.0	\pm7.6$
&\ \ $	 1390\pm77$
&\ \ $	 0.126	_{	-0.010	}^{+	0.007	}$
&\ \ $	 0.375	\pm0.089$
&\ \ $	 1.18	\pm0.14$
&\ \ $	 0.0457	\pm0.0033$
\\

$40<s<200$
&\ \ $	 87.4	\pm6.8$
&\ \ $	 1402	_{	-74	}^{+	75	}$
&\ \ $	 0.126	_{	-0.010	}^{+	0.007	}$
&\ \ $	 0.372	\pm0.080$
&\ \ $	 1.18	\pm0.13$
&\ \ $	 0.0454	\pm0.0028$
\\

South
&\ \ $	78.7	_{	-7.2	}^{+	6.7	}$
&\ \ $	1404	_{	-96	}^{+	98	}$
&\ \ $	0.137	_{	-0.016	}^{+	0.013	}$
&\ \ $	0.242	_{	-0.098	}^{+	0.092	}$
&\ \ $	1.45	_{	-0.26	}^{+	0.27	}$
&\ \ $	0.0401	_{	-0.0028	}^{+	0.0025	}$
\\

North
&\ \ $	92.5	_{	-8.8	}^{+	8.5	}$
&\ \ $	1371	_{	-85	}^{+	90	}$
&\ \ $	0.128	_{	-0.011	}^{+	0.008	}$
&\ \ $	0.414	\pm0.11$
&\ \ $	1.11	\pm0.15$
&\ \ $	0.0478	_{	-0.0038	}^{+	0.0037	}$
\\

300 mocks
&\ \ $	88.3	_{	-6.8	}^{+	6.6	}$
&\ \ $	1399	\pm74$
&\ \ $	0.129	_{	-0.010	}^{+	0.008	}$
&\ \ $	0.382	_{	-0.089	}^{+	0.090	}$
&\ \ $	1.19	\pm0.14$
&\ \ $	0.0457	_{	-0.0027	}^{+	0.0026	}$
\\
\hline
\end{tabular}

\begin{tabular}{crrrrrr}
\hline
 &$D_A/r_s$ &$D_V/r_s$ &$f\sigma_8$ &$A_s$ &$\alpha$ &$\epsilon	$ \\ \hline
 
fiducial result	
&\ \ $	 8.95	\pm0.26$
&\ \ $	 13.54	\pm0.28$
&\ \ $	 0.428	\pm0.066$
&\ \ $	 0.436	\pm0.017$
&\ \ $	 1.023	\pm0.021$
&\ \ $	 0.015	\pm0.027$\\

$25<s<160$
&\ \ $	 8.95	\pm0.26$
&\ \ $	 13.52	\pm0.28$
&\ \ $	 0.419	_{	-0.066	}^{+	0.068	}$
&\ \ $	 0.439	\pm0.012$
&\ \ $	 1.022	\pm0.021$
&\ \ $	 0.015	_{	-0.027	}^{+	0.028	}$
\\

$50<s<160$
&\ \ $	 8.99	\pm0.27$
&\ \ $	 13.56	\pm0.29$
&\ \ $	 0.419	_{	-0.071	}^{+	0.070	}$
&\ \ $	 0.438	\pm0.022$
&\ \ $	 1.025	\pm0.022$
&\ \ $	 0.012	\pm0.027$
\\

$40<s<120$
&\ \ $	 8.91	\pm0.30$
&\ \ $	 13.47	\pm0.31$
&\ \ $	 0.431	\pm0.072$
&\ \ $	 0.434	_{	-0.018	}^{+	0.019	}$
&\ \ $	 1.018	\pm0.024$
&\ \ $	 0.016	\pm0.032$
\\

$40<s<200$
&\ \ $	 8.98	\pm0.27$
&\ \ $	 13.57	\pm0.29$
&\ \ $	 0.433	_{	-0.065	}^{+	0.066	}$
&\ \ $	 0.437	\pm0.017$
&\ \ $	 1.026	\pm0.022$
&\ \ $	 0.015	\pm0.027$
\\

South
&\ \ $	9.17	_{	-0.32	}^{+	0.33	}$
&\ \ $	14.35	_{	-0.35	}^{+	0.40	}$
&\ \ $	0.33	_{	-0.11	}^{+	0.10	}$
&\ \ $	0.471	_{	-0.027	}^{+	0.028	}$
&\ \ $	1.085	_{	-0.027	}^{+	0.030   }$
&\ \ $	0.052\pm0.031$
\\

North
&\ \ $	8.82	_{	-0.33	}^{+	0.36	}$
&\ \ $	13.18	_{	-0.34	}^{+	0.37	}$
&\ \ $	0.448	_{	-0.078	}^{+	0.079	}$
&\ \ $	0.426	\pm0.020$
&\ \ $	0.996	_{	-0.025	}^{+	0.028	}$
&\ \ $	0.005	_{	-0.036	}^{+	0.034	}$
\\

300 mocks
&\ \ $	9.00	_{	-0.27	}^{+	0.26	}$
&\ \ $	13.57\pm0.27$
&\ \ $	0.443	_{	-0.070	}^{+	0.071	}$
&\ \ $	0.440\pm0.017$
&\ \ $	1.026\pm0.020$
&\ \ $	0.012	_{	-0.026	}^{+	0.027	}$
\\

\hline
\end{tabular}
\end{center}
\caption{This table presents the systematic tests with the scale range and the regime of the sample used.
The fiducial results are obtained by considering the scale range ($40<s<160\ h^{-1}$Mpc) from the combination of the north and south sample.
The other results are calculated with only specified quantities different from the fiducial one.
The unit of $H$ is $\Hunit$. The unit of $D_A$ and $r_s(z_d)$ is $\rm Mpc$.
} \label{table:test}
\end{table*}

\section{Summary} 
\label{sec:conclusion}
In this paper, we present measurements of the SDSS-III/BOSS DR9 anisotropic galaxy clustering and provide cosmological constraints from our CMASS data set only, and
in combination with other investigations in particular with CMB (WMAP9 and SPT) data.
We summarize our study as follows:
\\

(i) We present single-probe measurements of $H(0.57)$, $D_A(0.57)$, $f(0.57)\sigma_8(0.57)$, and $\Omega_mh^2$ as a 
summary of the information extracted from the BOSS CMASS galaxy clustering signal (results are listed in Table \ref{table:mean_fid} and \ref{table:covar_matrix_fid}).
A CosmoMC code that includes our BOSS CMASS clustering alone is provided\footnote{http://members.ift.uam-csic.es/chuang/BOSSDR9singleprobe}.
\\

(ii) Our cosmological constraints are obtained without assuming a flat universe or dark energy model. 
One can combine our results with other probe data sets and derive cosmological implications for a given model. 
We have explained the steps required to use our results and also demonstrated its applications with some examples.
\\

(iii) Combining our results from CMASS-only with CMB data, we find that the constraint on the constant equation of state of dark energy, $w$, 
can be significant improved by adding the measurement of $f(z)\sigma_8(z)$ to the $H(z)r_s$ and $D_A(z)/r_s$ measurements,
i.e., $f(z)\sigma_8(z)$ provides the strong power of the improvement going from spherical averaged analysis to the anisotropic galaxy clustering analysis.
Of course, the correlation among all the measurements must be taken into account; see \cite{Samushia:2012iq}.
Our results are all consistent with $w=-1$, which corresponds to the cosmological constant model (i.e. $\Lambda$CDM).
\\

(iv) We have compared our results with other investigations using the same BOSS DR9 CMASS data and find all the previous and new results are in excellent agreement.
We also found that our results are insensitive to the scale range used, which is likely due to the cancellation of the nonlinear effect and scale-dependent bias.
\\

(v) We explain and discuss the requirements of a given analysis to provide a single-probe measurement (see Sec. \ref{sec:single-probe}) in order to avoid double counting when combine with other cosmological probe data sets.
\\

Our methodology can be applied on the current and future large-scale galaxy surveys (e.g. eBOSS, BigBOSS, and Euclid) to obtain 
single-probe and model independent cosmological constraints, which will provide a powerful and convenient way to perform a joint data analysis with other data sets.

\section*{Acknowledgements}
We would like to thank Graeme Addison, Chris Blake, Ryan Keisler, Savvas Nesseris, Christian Reichardt, Beth Reid, Lado Samushia, and Kyle Story for useful discussions.
C.C. and F.P. acknowledge support from the Spanish MICINN’s Consolider-Ingenio 2010 Programme under grant MultiDark CSD2009-00064 and AYA2010-21231-C02-01 grant.
C.C. and F.P. were also supported by the Comunidad de Madrid under grant HEPHACOS S2009/ESP-1473

We acknowledge the use of the Legacy Archive for Microwave
Background Data Analysis (LAMBDA). Support
for LAMBDA is provided by the NASA Office of Space Science.
The mock catalogues used were produced in SCIAMA High Performance
Supercomputer (HPC) cluster, supported by the ICG, SEPNet and the
University of Portsmouth.

Funding for SDSS-III has been provided by the Alfred P. Sloan Foundation, the Participating Institutions, 
the National Science Foundation, and the U.S. Department of Energy Office of Science. The SDSS-III web 
site is http://www.sdss3.org/.

SDSS-III is managed by the Astrophysical Research Consortium for the Participating Institutions of the 
SDSS-III Collaboration including the University of Arizona, the Brazilian Participation Group, Brookhaven 
National Laboratory, University of Cambridge, Carnegie Mellon University, University of Florida, the French 
Participation Group, the German Participation Group, Harvard University, the Instituto de Astrofisica de 
Canarias, the Michigan State/Notre Dame/JINA Participation Group, Johns Hopkins University, Lawrence Berkeley 
National Laboratory, Max Planck Institute for Astrophysics, Max Planck Institute for Extraterrestrial 
Physics, New Mexico State University, New York University, Ohio State University, Pennsylvania State 
University, University of Portsmouth, Princeton University, the Spanish Participation Group, University of 
Tokyo, University of Utah, Vanderbilt University, University of Virginia, University of Washington, and Yale 
University.

\label{lastpage}

\end{document}